\documentclass[prb,showpacs,twocolumn,floats,10pt,aps,citeautoscript,longbibliography]{revtex4-2}

\usepackage{graphicx}
\usepackage{dcolumn}
\usepackage{bm}
\usepackage{lineno}
\usepackage{ulem}
\usepackage{soul}
\usepackage{color}
\usepackage{array}
\usepackage{booktabs}
\usepackage{multirow, makecell}
\usepackage{physics}
 \usepackage[colorlinks=true, allcolors=blue]{hyperref}%

\usepackage{amsmath}
\usepackage{enumitem}
\usepackage{algorithm}
\usepackage{algpseudocode}
\usepackage{comment}

\usepackage{booktabs} 
\usepackage{amssymb}
\usepackage{subfigure}
\usepackage{paralist} 
\usepackage{verbatim} 
\usepackage{exercise}
\usepackage[compat=1.1.0]{tikz-feynman}
\usepackage{framed}
\usepackage{amsthm}
\usepackage{natbib}  
\usepackage[toc,page]{appendix}

\usepackage{bm}
\usepackage{braket}
\usepackage{dcolumn}
\usepackage{esint}
\usepackage{mathrsfs}
\usepackage{mathtools}
\usepackage[version=3]{mhchem}

\allowdisplaybreaks
\renewcommand\vec[1]{\ensuremath\boldsymbol{#1}}

\renewcommand{\Re}{\operatorname{Re}}
\renewcommand{\Im}{\operatorname{Im}}
\renewcommand{\a}  {\alpha}    
\renewcommand{\b}  {\beta}     
\newcommand  {\g}  {\gamma}    
\renewcommand{\d}  {\delta}     \newcommand  {\D}  {\Delta}
\newcommand  {\e}  {\epsilon}  
\newcommand  {\z}  {\zeta}     
\renewcommand{\th} {\theta}     
\renewcommand{\l}  {\lambda}    
\newcommand  {\m}  {\mu}       
\newcommand  {\x}  {\xi}        
\newcommand  {\p}  {\pi}        
\newcommand  {\s}  {\sigma}     \renewcommand{\S}  {\Sigma}
\newcommand  {\ph} {\phi}       \newcommand  {\Ph} {\Phi}

\renewcommand{\o}  {\omega}     
\renewcommand {\dd} {\partial}
\renewcommand {\i} {\mathrm{i}}
\newcommand  {\up} {\uparrow}
\newcommand  {\dn} {\downarrow}
\newcommand  {\+}  {\dagger}
\newcommand{\etal}{\textit{et al.}}

\newcommand{\ourtitle}{External field induced metal-to-insulator transition in dissipative Hubbard model}

\begin{document}

\title{\ourtitle}

\author{Beomjoon Goh$^{1,2}$}
\author{Junwon Kim$^3$}
\author{Hongchul Choi$^1$}
\author{Ji Hoon Shim$^{1,3}$}
\email{jhshim@postech.ac.kr}

\affiliation{$^1$Department of Chemistry, Pohang University of Science and Technology, Pohang 37673, Korea} 
\affiliation{$^2$Institute for Data Innovation in Science, Seoul National University, Seoul 08826, Korea}
\affiliation{$^3$Division of Advanced Materials Science, Pohang University of Science and Technology, Pohang 37673, Korea}

\date{December 29, 2024}

\begin{abstract}
In this work, we develop a non-equilibrium steady-state non-crossing approximation (NESS-NCA) impurity solver applicable to general impurity problems.
The choice of the NCA as the impurity solver enables both a more accurate description of correlation effects with larger Coulomb interaction and scalability to multi-orbital systems.
Based on this development, we investigate strongly correlated non-equilibrium states of a dissipative lattice system under constant electric fields.
Both the electronic Coulomb interaction and the electric field are treated non-perturbatively using dynamical mean-field theory in its non-equilibrium steady-state form (NESS-DMFT) with the NESS-NCA impurity solver.
We validate our implementation using a half-filled single-band Hubbard model attached to a fictitious free Fermion reservoir, which prevents temperature divergence.
As a result, we identify metallic and insulating phases as functions of the electric field and the Coulomb interaction along with a phase coexistence region amid the metal-to-insulator transition (MIT).
We find that the MIT driven by the electric field is qualitatively similar to the equilibrium MIT as a function of temperature, differing from results in previous studies using the iterative perturbation theory (IPT) impurity solver.
Finally, we highlight the importance of the morphology of a correlated system under the influence of an electric field.
\end{abstract}

\maketitle

\section{Introduction}
Strongly correlated electron systems subjected to strong external electric fields can evolve into (quasi-) steady states over long timescales, exhibiting drastically different physics compared to equilibrium conditions.
Experimentally, interesting nonlinear transport phenomena such as oscillating current~\cite{Sawano2005}, negative differential resistance~\cite{Taguchi2000, Mori2009}, and colossal resistance~\cite{Asamitsu1997, Oshima1999, Liu2000} have been reported.
Field-induced phase transitions, driven by either photoexcitation or a dc field~\cite{Tokura2006, Yonemitsu2008}, where nonlinear optical responses can occur, have also been reported~\cite{Kishida2000, Mizuno2000}.
Furthermore, the interplay between Bloch oscillations induced by strong dc fields, field-driven electron localization (Wannier-Stark localization), and electronic interactions raises fascinating questions for further exploration.

These experimental discoveries have sparked growing interest in the theoretical understanding of out-of-equilibrium systems.
Multiple theoretical studies, supported by experiments, have investigated field-dependent phenomena in semiconductors including oxides~\cite{Lee2008, Driscoll2009, Morosan2012, Yuan2012, Lau2013}.
In particular, strong indications were found that the Joule heating mechanism occurs in binary oxides such as \ce{NiO} and \ce{VO_2} where correlated electrons might be responsible for the gap formation.
In these systems, electric-field-driven currents locally heat the material, resulting in temperature-driven resistive switching~\cite{Kim2007, Lee2008, Driscoll2009}.
These phenomena highlight the need for a unified theoretical framework that accounts for both electronic Coulomb interactions and external electric fields on a system, neither of which can be neglected.

Dynamical mean-field theory (DMFT)~\cite{Georges1996} has been one of the most successful methods for solving correlated lattice systems in equilibrium.
For systems driven out of equilibrium by relatively weak electric fields, linear response theory which considers the external field up to linear order from its equilibrium solution has proven effective.
However, as the strength of the electric field increases, the limitations of linear response theory become apparent.
For instance, in a Mott insulator, a strong electric field can cause dielectric breakdown, leading to a non-zero current, a phenomenon not captured by linear response theory.
To theoretically investigate such field-induced phenomena, non-equilibrium steady state dynamical mean-field theory (NESS-DMFT) with a suitable impurity solver provides a robust framework, which can treat both the electric field and the Coulomb interaction on equal footing.
Numerous insightful model studies using NESS-DMFT have reported results for non-equilibrium steady states under uniform static electric fields~\cite{Aron2012, Aron2013, Amaricci2012, Aoki2014, Li2015}.
The interplay between electronic interactions and external fields gives rise to a variety of exotic phenomena, demanding a comprehensive theoretical approach.

In this work, we develop a generalized NESS-DMFT framework capable of handling multi-orbital systems under various external fields and non-equilibrium steady state version of non-crossing approximation (NESS-NCA) impurity solver.
As a test case, we study the electronic structures of a single-band dissipative Hubbard model under static electric fields.
The NCA impurity solver is chosen for its superior ability to describe Coulomb correlation effects compared to previously tested solvers such as iterative perturbation theory (IPT) and its ease of generalization from equilibrium to multi-orbital non-equilibrium forms, 

This paper is organized as follows.
In Sec.~\ref{Model}, we introduce the dissipative Hubbard model.
Sec.~\ref{ssdmft} presents the formulation of the non-equilibrium steady-state dynamical mean-field theory for this model.
In Sec.~\ref{sec:Non-Crossing Approximation (NCA)}, we describe the non-crossing approximation impurity solver for non-equilibrium steady states.
Sec.~\ref{result} discusses the calculation results for the dissipative Hubbard model.
Finally, Sec.~\ref{conclude} concludes with our findings.

\section{Model}\label{Model}

\begin{figure}[t]
  \centering
  \includegraphics[width=0.7\linewidth]{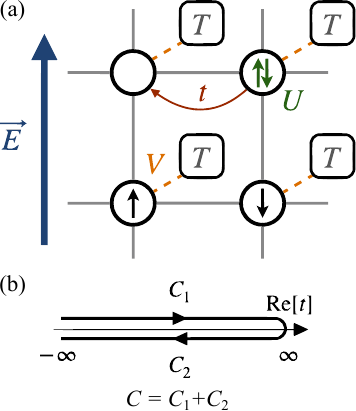}
  \caption{
  (a) The dissipative Hubbard model under a static and uniform electric field.
  Each correlated electron is coupled to a fermion with a coupling strength $V$.
  These fermions are assumed to be non-interacting and remain in equilibrium at a temperature $T_{B}$.}
  (b) The Keldysh contour.
  \label{fig:Model}
\end{figure}
We construct the single-particle Green’s function for a general time contour on the complex plane.
Using the contour-ordered Green’s function, we formulate the non-equilibrium version of dynamical mean-field theory.
The Hamiltonian is
\begin{align}\label{eq:hamiltonian}
  H =& \sum_{i} \mathbf{c}^{\+}_{i}(\bm{\e}+\bm{\ph}_i)\mathbf{c}_{i}
       -\sum_{\braket{ij}} [\mathbf{c}^{\+}_{i}\mathbf{t}^{ij}\mathbf{c}_{j} + \text{h.c}]
       +H_{U}(\mathbf{c}) \notag \\
     & +\sum_{ip} \mathbf{b}^{\+}_{ip}(\bm{\e}^{b}_{p}+\bm{\ph}_i)\mathbf{b}_{ip}
       - [\mathbf{c}^{\+}_{i}\mathbf{V}_p\mathbf{b}_{ip} + \text{h.c}]
\end{align}
where $\mathbf{c}^\+_{i} = [ c^{\+}_{i1}, c^{\+}_{i2}, \cdots, c^{\+}_{i\a}, \cdots, c^{\+}_{iN}]$, is a spinor of correlated fermions with $i$, and $\a$ being the site, basis (consists of orbitals, and spins) indices, respectively.
$\mathbf{b}_{ip}$ are free fermionic bath spinors with mode $p$, which are in equilibrium at temperature $T$, and are attached to each correlated fermion via an $N \times N$ coupling constant matrix $\mathbf{V}_{p}$~\cite{Tsuji2009, Amaricci2012, Aron2012, Werner2012, Han2013}.
The $N \times N$ diagonal matrices $\bm{\e} (\bm{\e}^{b}_{p}) and \bm{\ph}_i$ represents the energy levels and electric potential, respectively.
$\mathbf{t}^{ij}$ is the nearest neighbor hopping matrix elements in $N \times N$ block form.
The Coulomb gauge is used for the treatment of the electric field.
The electric potential for a static and uniform electric field in the Coulomb gauge is given by $[\bm{\ph}_i]_{\a\b} = - \left(\vec{r}_{i,\a}\cdot \vec{E}\right)\d_{\a\b}$, where $\vec{r}_{i,\a}$ represents a position vector of basis $\a$ at site $i$.
Fig.~\ref{fig:Model}(a) illustrates this dissipative Hubbard model under an electric field.
After integrating out the free fermion reservoir (heat bath), the action becomes
\begin{align}\label{eq:action}
  \mathcal{S} = -\i\int_\mathcal{C} dt & \Big\{
    \sum_{i} \mathbf{c}^{*}_{i}(t) \left[(\i\dd_t +\m)\mathbf{I} -\bm{\e} -\bm{\ph}_i \right] \mathbf{c}_{i}(t) \notag \\
    & +\sum_{\braket{ij}} \left[\mathbf{c}^{*}_{i}(t) \mathbf{t}^{ij} \mathbf{c}_{j}(t) + \text{h.c} \right] -H_{U}[\mathbf{c}(t)] \notag \\
    & +\int_\mathcal{C} dt' \sum_{i} \mathbf{c}^{*}_{i}(t) \bm{\S}_{\Gamma i}(t,t') \mathbf{c}_{i}(t') 
     \Big\}
\end{align}
where,
\begin{equation}
  \bm{\S}_{\Gamma i}(t,t') = \sum_{p}\mathbf{V}_{p}^{\+}\mathbf{G}^{\Gamma}_{i,p}(t,t')\mathbf{V}_{p},
\end{equation}
is the reservoir self-energy coming from the coupling to the free fermions in equilibrium and $\mathbf{G}^{\Gamma}_{i,p}(t,t') = -\i\braket{T_{\mathcal{C}} \mathbf{b}_{ip}(t) \mathbf{b}^{\+}_{ip}(t')}_{\text{bath}}$.
Now, an assumption is made: $\mathbf{t}^{ij}$ is separable along the direction of $\vec{E}$.
Perpendicular to direction the non interacting Hamiltonian can be diagonalized with $k_{\perp}$ as
\begin{align}
  H^0 = & \sum_{l,k_{\perp}} \mathbf{c}^{\+}_{l,k_{\perp}} (\bm{\e}+\bm{\ph}_l+\bm{\e}_{k_{\perp}}) \mathbf{c}_{l,k_{\perp}} \notag \\
   +&\sum_{l,k_{\perp}} [\mathbf{c}^{\+}_{l,k_{\perp}}\mathbf{t}^{l,l+1} \mathbf{c}_{l+1,k_{\perp}} + \text{h.c} ]
\end{align}
where $l$ is the site index parallel to the field direction.

The following local action can be defined by the DMFT formulation
\begin{equation}\label{eq:action2}
  \mathcal{S}_{\text{loc}}[\mathbf{c}]
    = \int_\mathcal{C} dt H_{\text{loc}}(t) +\int_\mathcal{C} dtdt' \mathbf{c}^{*}(t)\bm{\D}(t,t')\mathbf{c}(t').
\end{equation}
This impurity problem is solved by the non-equilibrium NCA on the Keldysh contour (Sec.~\ref{sec:Non-Crossing Approximation (NCA)}).

\section{NESS-DMFT}\label{ssdmft}

DMFT for non-equilibrium can be formally constructed in the same way as for the equilibrium case using the contour-ordered formalism~\cite{Aoki2014}.
The lattice Green's function with the DMFT local self-energy is
\begin{equation}\label{eq:Glatt}
  [G_{ij}(t,t')]^{-1} = [(\i\dd_t + \m)\d_{ij} - h_{ij}(t)]\d_{\mathcal{C}}(t,t') - \S_{ii}(t,t')
\end{equation}
where $h_{ij}(t)$ is the lattice Hamiltonian that can have explicit time dependency.
The local Green's function $G_{ii}(t,t')$ is then equated to an impurity Green's function $G_{\text{imp}}(t,t')$ of the impurity action comprised of the local Hamiltonian and the self-consistently determined hybridization function $\D(t,t')$ so that
\begin{equation}
  [G_{ii}(t,t')]^{-1} = (\i\dd_t + \m - h_{ii}(t))\d_{\mathcal{C}}(t,t') - \D(t,t') - \S_{ii}(t,t')
\end{equation}
If the Hamiltonian does not have explicit time dependency but have an external field that drives the system out of equilibrium into a steady state, only the time difference matters.
The Keldysh contour (see Fig.~\ref{fig:Model}(b)) is most suitable for such scenarios as it exploits the time translational symmetry.
Contour-ordered functions need to be analytically continued to real and/or imaginary time axs, procedure often referred to as the Langreth-Wilkins rules~\cite{Langreth:1972}.
By applying these rules, the Dyson's equation on the Keldysh contour can be written with physical (retarded and lesser) Green's function components
\begin{align}
  [\i\dd_t-h] G^{R}(t)
    & - \int dt' \S^{R}(t-t')G^{R}(t') = \d(t) \\
  [\i\dd_t-h] G^{<}(t) 
    & - \int dt' \S^{R}(t-t')G^{<}(t') \notag \\
    & - \int dt' \S^{<}(t-t')G^{A}(t') = 0.
\end{align}
After the Fourier transformation to the real energy axis, 
\begin{align}
  G^{R}(\o) & = \left[ \o-h-\S^{R}(\o)\right]^{-1} \\
  G^{<}(\o) & = G^{R}(\o)\S^{<}(\o)G^{A}(\o).
\end{align}

We require an expression for the local Green’s function derived from the lattice Green’s function (Eq.~\ref{eq:Glatt}).
In equilibrium, this procedure is usually straightforward, since the lattice Green’s function is diagonalizable in momentum space.
With external field written in Coulomb gauge, however, potential differences along the field direction impede the standard Fourier transformation.
Following the work of Li \etal~\cite{Li2015} the local Green’s function can instead be obtained via an iterative procedure.

\subsection{Local Green's function}

Since we have the DMFT self-energy, the retarded component of the Green's function for the above action (Eq.~\ref{eq:action}) can be written explicitly in tri-diagonal block matrix form.
\begin{equation}\label{eq:GRdef}
  \left[G^{R}(k_\perp, \o)\right]^{-1} = 
    \begin{bmatrix}
      \ddots & \ddots & & & \\
      \ddots & \mathbf{A}_{-1} & \mathbf{t} & & \\
      & \mathbf{t} & \mathbf{A}_{0} & \mathbf{t} & \\
      & & \mathbf{t} & \mathbf{A}_{1} & \ddots \\
      & & & \ddots & \ddots \\
    \end{bmatrix},
\end{equation}
where we define $\mathbf{A}_{l} \equiv (\o+\m-\ph_l)\mathbf{I} -\bm{\e} -\bm{\e}_{k_{\perp}} - \bm{\S}^{R}_{\text{tot}}(\o-\ph_l)$, $\mathbf{t} \equiv \mathbf{t}^{l,l+1}$, and the total self-energy $\bm{\S}_{\text{tot}}(\o) = \bm{\S}_{\Gamma}(\o) + \bm{\S}_{U}(\o)$.
We define, for convenience, a self-similar function
\begin{equation}\label{eq:FRdef}
    \mathbf{F}^{R}_{\pm}(k_\perp, \o-\ph_{\pm1}) =
    \left[\mathbf{A}_{\pm1}-\mathbf{t}\left[\mathbf{A}_{\pm2}-\cdots
    \mathbf{t}\right]^{-1}\mathbf{t}\right]^{-1}
\end{equation}
The retarded component of the local Green's function is $\mathbf{G}^{R}(\o) = \sum_{k_\perp} \mathbf{G}^{R}(k_\perp, \o)_{00}$, where,
\begin{align}
  \left[\mathbf{G}^{R}(k_\perp, \o)_{00}\right]^{-1}
  = \mathbf{A}_{0} - &\mathbf{t}
      \big( \mathbf{F}^{R}_{+}(k_\perp, \o-\ph_1) \notag \\
      & + \mathbf{F}^{R}_{-}(k_\perp, \o-\ph_{-1}) \big) \mathbf{t}.
\end{align}

Using the Dyson's equation of the lesser component of the action (Eq.~\ref{eq:action}),
\begin{equation}\label{eq:DysonG<}
  \mathbf{G}^{<}(k_\perp,\o)_{ll'}
  = \sum_{p}\mathbf{G}^{R}(k_\perp,\o)_{lp}\bm{\S}^{<}_{\text{tot}}(\o)_{p} \mathbf{G}^{A}(k_\perp,\o)_{pl'}
\end{equation}
since the self-energy is block diagonal by the virtue of DMFT.
By a similar procedure we get the lesser component of the local Green's function $\mathbf{G}^{<}(\o) = \sum_{k_\perp} \mathbf{G}^{<}(k_\perp, \o)_{00}$.
\begin{align}
  \mathbf{G}^{<}(k_\perp, \o)_{00}
  &=
  \mathbf{G}^{R}(k_\perp, \o)_{00}
    \big[
      \bm{\S}^{<}_{\text{tot}}(\o) + \mathbf{t}
      \big(
        \mathbf{F}^{<}_{+}(k_\perp,\o-\ph_{1})  \notag \\
        & +\mathbf{F}^{<}_{-}(k_\perp,\o-\ph_{-1})
      \big)
      \mathbf{t}
    \big] \mathbf{G}^{A}(k_\perp, \o)_{00}
\end{align}
where $\mathbf{F}^{<}$ is another self-similar function defined as,
\begin{align}\label{eq:F<recur}
  \mathbf{F}^{<}_{\pm}(k_\perp,\o&-\ph_{\pm1})
  = \mathbf{F}^{R}_{\pm}(k_\perp,\o-\ph_{\pm1}) \big[
      \bm{\S}^{<}_{\text{tot}}(\o-\ph_{\pm1}) \notag \\ 
     & +\mathbf{t} \mathbf{F}^{<}_{\pm}(k_\perp,\o-\ph_{\pm2}) \mathbf{t}
    \big] \mathbf{F}^{A}_{\pm}(k_\perp,\o-\ph_{\pm1})
\end{align}

\subsection{Heat reservoir self-energy}

If one uses a constant heat reservoir self-energy $\S_{\Gamma\a}(\o) = i\Gamma$, as done previously by many~\cite{Li2015}, the current formulation of NCA fails numerically.
This failure arises because the reservoir self-energy is incorporated into the hybridization function, resulting in a fictitious non-zero value at large $\pm\o$.
In the impurity solver, the hybridization functions are convoluted with various pseudoparticle Green's functions to obtain the pseudoparticle self-energy.
The resulting imaginary part of the self-energy also has a non-zero value at large $\pm\o$.
However, in our formalism the real part of the retarded self-energy is calculated using the Kramers-Kronig relation and one of whose key assumptions is that the function in question must vanish at large $\pm\o$.
To avoid this issue, We employ the following form of the reservoir self-energy:
\begin{equation}
  \S^{R}_{\Gamma\a}(\o) = -i\Gamma w(\frac{\o}{\sqrt{\p}W_\Gamma}).
\end{equation}
where, $w(x) = \text{erfc}(-ix)e^{-x^2}$ is the Faddeeva function.
The lesser component can be obtained using the fluctuation-dissipation theorem, since the reservoir is in equilibrium at the heat bath temperature $T_{B} = 1/\b_{B}$.
\begin{equation}
  \Im{\S^{<}_{\Gamma\a}(\o)} = -\frac{2}{e^{\b_{B}\o}+1} \Im{\S^{<}_{\Gamma\a}(\o)}
\end{equation}
In this form of the reservoir self-energy, as long as the energy mesh is sufficiently wide for the $W_{\Gamma}$, the input hybridization functions becomes zero at large $\pm\o$ thereby satisfying the necessary condition for the Kramers-Kornig realation.

\subsection{Hybridization function}

From the impurity action (Eq.~\ref{eq:action2}) one can write the Dyson's equation and get the hybridization functions.
\begin{align}
  \mathbf{G}^{R}(\o) &=
    \left[
      (\o+\m)\mathbf{I} - \bm{\e} - \bm{\D}^{R}(\o) - \bm{\S}^{R}_{U}(\o)
    \right]^{-1} \\
  \mathbf{G}^{<}(\o) &=
    \mathbf{G}^{R}(\o)
    \left[
      \bm{\D}^{<}(\o) + \bm{\S}^{<}_{U}(\o)
    \right]
  \mathbf{G}^{A}(\o)
\end{align}
Inverting these,
\begin{align}
  \bm{\D}^{R}(\o) &=
    (\o+\m)\mathbf{I} - \bm{\e} - \left[ \mathbf{G}^{R}(\o) \right]^{-1} - \bm{\S}^{R}_{U}(\o) \\
  \bm{\D}^{<}(\o) &=
    \left[ \mathbf{G}^{R}(\o) \right]^{-1} \mathbf{G}^{<}(\o) \left[ \mathbf{G}^{A}(\o) \right]^{-1}
    - \bm{\S}^{<}_{U}(\o)
\end{align}
The information of the heat reservoir self-energy, the heat bath temperature $T_{B}$ and the dissipation rate $\Gamma$ is implicitly contained in the hybridization function.
The NESS-NCA will calculate new local Green's function with the given hybridization function and update the local self-energy of the system.

\section{NESS-NCA}
  \label{sec:Non-Crossing Approximation (NCA)}
The NCA is the lowest order pertubative approximation for the self-energy in the strong coupling limit.
We choose NCA since it is believed to offer a more accurate description of strongly correlated systems with larger Coulomb interactions than the previously employed iterative perturbation theory (IPT), and it can be readily extended to multi-orbital systems.
The effective impurity action of the type
\begin{align} \label{eq:effImpurity}
  S_{\text{loc}}
    =& \int_\mathcal{C} dt \sum_{\a} c^*_\a(t)(\dd_t-\m) c_\a(t) - H_{\text{loc}}[c^{*}(t),c(t)] \notag \\
    &+ \int_\mathcal{C} dt \int_\mathcal{C} dt' \sum_{\a\b}
        c^*_\a(t)\D_{\a\b}(t,t')c_\b(t') 
\end{align} 
is to be solved by the NCA, where $\a$, $\b$ are spin and orbital indices and $\mathcal{C}$ represents a general contour on complex time domain.
The single impurity Anderson model is described by $H_{\text{loc}} = \sum_\s \e n_{\s} -U n_{\up}n_{\dn}$.
The resummation of the diagrams to infinite order is problematic since the unperturbed system ($U=0$) is not quadratic.
One technique to circumvent this problem is the so called pseudoparticle approach~\cite{Coleman1984,Bickers1987,Wingreen1994,Haule2001,Eckstein2010}.
It introduces a set of pseudoparticles, hence expanding the Hilbert space, to make the local action quadratic in terms of those pseudoparticles.
In return, one has to project the observable to the physical Hilbert space to have physically meaningful one.

\subsection{Pseudoparticle representation}

The local part of the impurity Hamiltonian $H_\text{loc}$ is assumed to be represented in a local basis set
$\ket{m}$
\begin{equation}
  H_\text{loc} = \sum_{mm'} \ket{m}h_{mm'}\bra{m'}.
\end{equation}
For example, in the case of single Anderson impurity, one might take $\ket{m}$ as $\ket{0}, \ket{\up}, \ket{\dn}$, and $\ket{\up\dn}$.
For each state $\ket{m}$ one pseudoparticle field $a_m^{(*)}$ is introduced whose statistics is bosonic (fermionic) if the state has an even (odd) number of particles.
Note that this includes even the vacuum state $\ket{0}$ so that a new vacuum is introduced as well, $\ket{vac}$.
In this way one has an isomorphism $\ket{m} \leftrightarrow a_m^{\+}\ket{vac}$ which expands the original Hilbert space to the pseudoparticle Fock space $F_Q$.
The physical Hilbert space can be identified with the Fock space in which the total number of pseudoparticles $Q = \sum_m a_m^* a_m$ is equal to the total number of physical electrons $N$.
And any observable $A$ can be computed in the pseudoparticle Fock space and then be projected onto the $Q = N$ subspace assuming the pseudoparticle action and the observable is constructed in such a way that they are equal to physical ones in the $Q=N$ subspace.
This can be done by defining

\begin{equation}
  c_\a^* = \sum_{mn} F^{\a*}_{mn} a_m^* a_n, \quad c_\a = \sum_{mn} F^{\a}_{nm} a_m^* a_n
\end{equation}
where $F^{\a}_{mn} $ is the annihilation operator matrix element of $\braket{m|c_\a|n}$ for pseudoparticle  indices $m,n$ and the local fermion indices $\a,\b$.
Note that we use roman subscripts ($m,n,\ldots$) for pseudoparticle quantities, and Greek characters ($\a,\b,\ldots$) for local, physical quantities.
  
Then the action becomes
\begin{align}
  \mathcal{S}_{\text{loc}}[a]
  &= \int_C dt \sum_{mn} a_m^*(t) [i\dd_t-h_{mn}(t)]a_{n}(t) \notag\\
  +& \int_C dtdt' \sum_{\text{all}}
      a_m^*(t)a_{m'}(t) \notag \\
      &\times \Big[ F^{\b*}_{mm'}\D_{\b\a}(t,t')F^{\a}_{n'n} \Big]
      a_{n}^*(t')a_{n'}(t').
\end{align}
To project out the subspace $F_1$, a chemical potential $-\l$ which is associated with the charge $Q\in\mathbb{Z}$ is introduced to have a grand canonical partition function that is the sum of the canonical partition function for each $F_Q$.
Defining also the associated fugacity $\z=e^{-\b\l}$
\begin{equation}
  \mathcal{Z}_G = \text{Tr}(e^{-\b(H+\l Q)}) = \sum_{Q=0}^{\infty} \mathcal{Z}_{Q} e^{-\b\l Q},
\end{equation}
projection is carried out as
\begin{equation}
  \mathcal{Z}_{1} = \lim_{\l\to\infty} \frac{\dd}{\dd\z}\mathcal{Z}_G.
\end{equation}
Since any observable annihilate for the $Q=0$ state, the physical observable $\braket{A}$ are projected as $\lambda$ goes to infinity.
\begin{equation}
  \braket{A} = \lim_{\l\to\infty} \frac{\braket{QA}_G}{\braket{Q}_G}
\end{equation}

Working on the grand canonical ensemble, the usual diagrammatic perturbation technique can be used with respect to the local (atomic) Hamiltonian.
From the grand canonical ensembles with superscript $\l$,
we denote the dressed pseudoparticle Green's function, pseudoparticle self-energy, local Green's function, and the total number of pseudoparticle number as
$G^{\l}_{mn}(t,t')$, $\S^{\l}_{mn}(t,t')$, $G^{\l}_{\a\b}(t,t')$ and $Q^{\l}$, respectively.
We identify the pseudoparticle Green's function and the four point interaction vertex as
\begin{equation}
\begin{gathered}
  \begin{tikzpicture}
    \begin{feynman}
      \vertex (a0) {$m$};
      \vertex [right=2.5cm of a0] (b0) {$n,$};
      \diagram* {
        (b0) -- [fermion, edge label'= $G^{\l}_{mn}(t{,}t')$] (a0)
      };
      \vertex [below=0.7cm of a0] (dum) {};
    \end{feynman}
  \end{tikzpicture}
  \quad
  \begin{tikzpicture}
    \begin{feynman}
      \vertex (a0); \vertex [left=0.2cm of a0] (aa0) {$F^{\b*}_{mm'}$};
      \vertex [right=of a0] (b0); \vertex [right=0.2cm of b0] (aa0) {$F^{\a}_{n'n}$};
      \vertex [below left=0.5cm of a0] (i1) {$m'$};
      \vertex [above left=0.5cm of a0] (f1) {$m$};
      \vertex [below right=0.5cm of b0] (i2) {$n$};
      \vertex [above right=0.5cm of b0] (f2) {$n'$};
      \diagram* {
        (i1) -- [fermion] (a0)
             -- [fermion] (f1),
        (b0) -- [charged scalar, edge label'= $\D_{\b\a}(t{,}t')$] (a0),
        (i2) -- [fermion] (b0)
             -- [fermion] (f2),
      };
    \end{feynman}
  \end{tikzpicture}
\end{gathered}
\label{eq:NCAFeynman}
\end{equation}
The self-energy expansion of the full Green's function gives the usual Dyson's equations.

The Luttinger-Ward functional in the first two order is 
\begin{equation}
  \Ph[G^{\l},\D] =
  \begin{gathered}
  \begin{tikzpicture}
    \begin{feynman}[large]
      \vertex (a0);
      \vertex [right=of a0] (a1);
      \diagram* {
        (a0) -- [fermion, half left, looseness = 1.65] (a1)
             -- [fermion, half left, looseness = 1.65] (a0),
        (a0) -- [scalar] (a1),
      };
      \vertex [right=0.2cm of a1] (pl){$+$};
    \end{feynman}
    \begin{feynman}[small]
      \vertex [right=0.4cm of pl] (b1);
      \vertex [right=of b1] (O0);
      \vertex [above=of O0] (b2);
      \vertex [right=of O0] (b3);
      \vertex [below=of O0] (b4);
      \diagram* {
        (b1) -- [fermion, quarter left] (b2)
             -- [fermion, quarter left] (b3)
             -- [fermion, quarter left] (b4)
             -- [fermion, quarter left] (b1),
        (b1) -- [scalar] (b3),
        (b2) -- [scalar] (b4),
      };
    \end{feynman}
  \end{tikzpicture}
  \end{gathered}
  \label{eq:NCALuttinger}
\end{equation}
Differentiating the first diagram of Eq.~\ref{eq:NCALuttinger} with respect to the pseudo Green's function, we get the lowest order self-energy which has two diagrams
\begin{equation}
  \S^{\l}_{mn}(t,t') = \ 
  \begin{gathered}
  \begin{tikzpicture}
    \begin{feynman}
      \vertex (a0);
      \vertex [right=0.4cm of a0] (a1);
      \vertex [right=of a1] (a2);
      \vertex [right=0.4cm of a2] (a3);
      \vertex [above right=0.2cm of a3] (pl) {$+$};
      \vertex [below right=0.5cm of pl] (b0);
      \vertex [right=0.4cm of b0] (b1);
      \vertex [right=of b1] (b2);
      \vertex [right=0.4cm of b2] (b3);
      \diagram* {
        (a0) -- [fermion] (a1) -- [fermion] (a2) -- [charged scalar, half right, looseness = 1.65] (a1),
        (a2) -- [fermion] (a3),
        (b0) -- [fermion] (b1) -- [fermion] (b2) -- [anti charged scalar, half right, looseness = 1.65] (b1),
        (b2) -- [fermion] (b3),
      };
    \end{feynman}
  \end{tikzpicture}
  \end{gathered}
\label{eq:NCAselfenergy}
\end{equation}
The sign of each diagram is $-1^{s+f}$, where $s$ is number of crossings of hybridization lines, and $f$ is the number of hybridization lines that go opposite direction to the pseudoparticle lines.

The local Green's function is obtained by differentiating the functional with respect to the hybridization function.
By truncating the functional at the same order, one gets conserving approximation.
The lowest order local Green's function in the grand canonical ensemble is then,
\begin{equation}
  G^{\l}_{\a\b}(t,t') =
  \begin{gathered}
  \begin{tikzpicture}
    \begin{feynman}
      \vertex (a0);
      \vertex [right=0.7cm of a0] (a1);
      \vertex [right=of a1] (a2);
      \vertex [right=0.7cm of a2] (a3);
      \diagram* {
        (a0) -- [charged scalar] (a1),
        (a1) -- [fermion, half left, looseness = 1.65] (a2)
             -- [fermion, half left, looseness = 1.65] (a1),
        (a2) -- [charged scalar] (a3),
      };
    \end{feynman}
  \end{tikzpicture}
  \end{gathered}
\label{eq:NCAlocalGreen's}
\end{equation}
The vertex corrections, i.e., one crossing approximation (OCA) or higher orders, result in a certain number of closed loops of pseudoparticles.
Each closed loop is $\mathcal{O}(e^{-\b\l})$ so that in $\l\to\infty$ limit any diagram that has loops vanish relative to the lowest order with $Q=1$.

\subsection{Pseudoparticle quantities}

Since the contour chosen is the Keldysh contour, Fourier transform from time to energy is possible so that the retarded and the lesser components of dressed pseudoparticle Green's functions are
\begin{align}
  [G^{R,\l}(\o)]^{-1}_{mn}
    &= \o - h_{mn} - \S^{R,\l}_{mn}(\o)
    \label{eq:CompGRmnL} \\
  G^{<,\l}_{mn}(\o)
    &= \sum_{pq} G^{R,\l}_{mp}(\o) \S^{<,\l}_{pq}(\o) G^{A,\l}_{qn}(\o)
    \label{eq:CompG<mnL}
\end{align}

The pseudoparticle self energies, Eq.~\ref{eq:NCAselfenergy}, are explicitly
\begin{align}
  \S^{\l}_{mn}(t,t') = 
    i& \sum_{\a\b m'n'} -F^{\b}_{mm'}F^{\a*}_{n'n} G^{\l}_{m'n'}(t,t') \D_{\a\b}(t',t)
    \notag \\
    &+ F^{\b*}_{mm'}F^{\a}_{n'n} G^{\l}_{m'n'}(t,t') \D_{\b\a}(t,t')
\end{align}
Analytic continuation from contour-ordered function to retarded function leads to
\begin{align}
  \S^{R,\l}_{mn}(t,t') =&
    i\sum_{\a\b m'n'}
     -F^{\b}_{mm'}F^{\a*}_{n'n}
      \Big[
         G^{R,\l}_{m'n'}(t,t') \D^{<}_{\a\b}(t',t)
        \notag \\
        &+G^{<,\l}_{m'n'}(t,t') \D^{A}_{\a\b}(t',t)
      \Big]
      \notag \\
  &   +F^{\b*}_{mm'}F^{\a}_{n'n}
      \Big[
         G^{R,\l}_{m'n'}(t,t') \D^{>}_{\b\a}(t,t')
        \notag \\
        &+G^{<,\l}_{m'n'}(t,t') \D^{R}_{\b\a}(t,t')
      \Big]
\end{align}
There is time translational symmetry in Keldysh formalism so only the time difference matters $(t,t') \to t-t'$ and the reference time can be set to zero $t'=0$.
The retarded self-energy is then
\begin{align}
  \S^{R,\l}_{mn}(\o) &=
    i\sum_{\a\b m'n'}\int\frac{d\x}{2\p}
    \Big\{ \notag \\
  &   -F^{\b}_{mm'}F^{\a*}_{n'n}
      \Big[
         G^{R,\l}_{m'n'}(\x) \D^{<}_{\a\b}(\x-\o)
      \notag \\                                 
      &\qquad\qquad\quad                       
        +G^{<,\l}_{m'n'}(\x) \D^{A}_{\a\b}(\x-\o)
      \Big]
      \notag \\
  &   +F^{\b*}_{mm'}F^{\a}_{n'n}
      \Big[
         G^{R,\l}_{m'n'}(\x) \D^{>}_{\b\a}(\o-\x)
      \notag \\
      &\qquad\qquad\quad
        +G^{<,\l}_{m'n'}(\x) \D^{R}_{\b\a}(\o-\x)
      \Big]
    \Big\}
    \label{eq:CompSRmnL}
\end{align}
where we used the convolution theorem:
\begin{equation}
  \mathcal{F}[f*g] = \mathcal{F}[f]\cdot\mathcal{F}[f], \quad
  \mathcal{F}[f\cdot g] = \mathcal{F}[f]*\mathcal{F}[f]
\end{equation}
with point-wise multiplication $\cdot$ and convolution $*$.
Likewise, the lesser self-energy is
\begin{align}
  \S^{<,\l}_{mn}(t,t')
  =&
    i\sum_{\a\b m'n'}
    \Big\{
      -F^{\b}_{mm'}F^{\a*}_{n'n} G^{<,\l}_{m'n'}(t,t') \D^{>}_{\a\b}(t',t)
      \notag \\
      &+F^{\b*}_{mm'}F^{\a}_{n'n} G^{<,\l}_{m'n'}(t,t') \D^{<}_{\b\a}(t,t')
    \Big\} \\
  \S^{<,\l}_{mn}(\o)
  =&
    i\sum_{\a\b m'n'}\int\frac{d\x}{2\p}
    \Big\{
      \notag \\
      &-F^{\b}_{mm'}F^{\a*}_{n'n} G^{<,\l}_{m'n'}(\x) \D^{>}_{\a\b}(\x-\o)
      \notag \\
      &+F^{\b*}_{mm'}F^{\a}_{n'n} G^{<,\l}_{m'n'}(\x) \D^{<}_{\b\a}(\o-\x)
    \Big\}
    \label{eq:CompS<mnL}
\end{align}
  
Lastly, The total pseudoparticle number is
\begin{align}
  Q^{\l} 
  &= \sum_{m}\Braket{a^{\+}_{m}(t'=0^{+})a_{m}(0)}_{\l} \notag\\
  &= \sum_{m} -i^{2} (-1)^{2m}\Braket{a^{\+}_{m}(t'=0^{+})a_{m}(0)}_{\l}
\end{align}
where, $0^{+}$ comes infinitesimally later than $t=0$ on the contour and $(-1)^{m} = 1(-1)$ if the statistics is bosonic(fermionic).
By the definition of the lesser Green's function and the Fourier transformation, $Q^{\l}$ becomes
\begin{align}
  Q^{\l}
  &= i\sum_{m} (-1)^{m} G^{<,\l}_{mm}(t=0^{-}) \notag\\
  &= i\sum_{m} (-1)^{m} \int \frac{d\x}{2\p} e^{-i\x(0^{-})} G^{<,\l}_{mm}(\x) \notag\\
  &= i\sum_{m} (-1)^{m} \int \frac{d\x}{2\p} G^{<,\l}_{mm}(\x)
\end{align}

Following the work of Eckstein~\cite{Eckstein2010}, the projection of pseudoparticles quantities is 
\begin{align}
  \lim_{\l\to\infty} &G^{\l}(t,t')  \notag \\
  =& \lim_{\l\to\infty}
      \big[
        e^{\l\Im(t-t')} (\th_{\mathcal{C}}(t,t') + \th_{\mathcal{C}}(t',t)e^{-\b\l})
      \big] \notag \\
      \quad&\times \left(G(t,t') + \mathcal{O}(e^{-\b\l}) \right)
      \notag \\
  =& \lim_{\l\to\infty} \big\{
        \th_{\mathcal{C}}(t,t') \left[G(t,t') + \mathcal{O}(e^{-\b\l}) \right]
      \notag \\
      \quad&+\th_{\mathcal{C}}(t',t)e^{-\b\l}G(t,t') \big\} \notag \\
  =& \th_{\mathcal{C}}(t,t') G(t,t') + \th_{\mathcal{C}}(t',t)G(t,t') \lim_{\l\to\infty} e^{-\b\l}
\end{align}
Hence, if $t$ comes later than $t'$ on the contour,
\begin{equation}
  G^{>}(t,t') = \lim_{\l\to\infty} G^{>,\l}(t,t')
\end{equation}
and if $t'$ comes later than $t$,
\begin{equation}
  G^{<}(t,t') = \lim_{\l\to\infty} e^{\b\l} G^{<,\l}(t,t')
\end{equation}
so that the retarded component is
\begin{align}
  \lim_{\l\to\infty} G^{R,\l}(t,t')
  =& \lim_{\l\to\infty} \th(t-t') [ G^{>,\l}(t,t') - G^{<,\l}(t,t') ] \notag \\
  =& \th(t-t')G^{>}(t,t') \notag \\
  =& \th_{\mathcal{C}}(t,t') G^{R}(t,t').
\end{align}
Note that the lesser component vanishes after taking the limit due to the factor $\mathcal{O}(e^{-\b\l})$.
For $\th_{\mathcal{C}}(t,t') = 1$ case $G^{R}(t) =  \th(t)G^{>}(t)$ and Fourier transform results in
\begin{equation}
  G^{R}(\o) = \int \frac{d\x}{2\p} \frac{iG^{>}(\x)}{\x-\o} + \p\d(\o-\x)G^{>}(\x)
\end{equation}
since $\mathcal{F}[\th(t)] = 1/i\o + \p\d(\o)$.

After the projection, the Dyson's equation and thus the functional form of pseudo Green's function do not change:
\begin{align}
  [G^{R}(\o)]^{-1}_{mn}
    &= \o - h_{mn} - \S^{R}_{mn}(\o)
    \label{eq:CompGRmn} \\
  G^{<}_{mn}(\o)
    &= \sum_{pq} G^{R}_{mp}(\o) \S^{<}_{pq}(\o) G^{A}_{qn}(\o)
    \label{eq:CompG<mn}
\end{align}
However, the self-energy simplifies to
\begin{align}
  &\S^{R,<}_{mn}(\o) =
    -i\sum_{\a\b m'n'}\int\frac{d\x}{2\p}G^{R,<}_{m'n'}(\x) \notag \\
    &\ \times\big\{F^{\b}_{mm'}F^{\a*}_{n'n} \D^{<,>}_{\a\b}(\x-\o) -F^{\b*}_{mm'}F^{\a}_{n'n} \D^{>,<}_{\b\a}(\o-\x) \big\}
    \label{eq:CompSmn}
\end{align}
Hence, we have a closed set of equations for the pseudoparticles.
Starting with a guess self-energy, build up Green's function (Eq.~\ref{eq:CompGRmn},~\ref{eq:CompG<mn}), calculate the self-energy (Eq.~\ref{eq:CompSmn}) which will be fed to the Green's function iteratively until self consistency is met.

\subsection{Local Green's function}

The local Green's function, Eq.~\ref{eq:NCAlocalGreen's}, is obtained by taking functional derivative of the lowest order Luttinger-Ward functional with respect to the hybridization function.
\begin{equation}
  G^{\l}_{\a\b}(t,t') = i\sum_{mm'nn'} F^{\b*}_{mm'}F^{\a}_{n'n} G^{\l}_{nm}(t,t') G^{\l}_{m'n'}(t',t) 
\end{equation}
Exactly same procedure leads to
\begin{align}
  G^{R,\l}_{\a\b}(\o)
  =&  i\sum_{mnm'n'} F^{\b*}_{mm'}F^{\a}_{n'n} \int \frac{d\x}{2\p}
        G^{R,\l}_{nm}(\x) G^{<,\l}_{m'n'}(\x-\o)  \notag \\
       &+G^{<,\l}_{nm}(\x) G^{A,\l}_{m'n'}(\x-\o) 
  \\
  G^{<,\l}_{\a\b}(\o)
  =&  i\sum_{mnm'n'} F^{\b*}_{mm'}F^{\a}_{n'n} \int \frac{d\x}{2\p}
        G^{<,\l}_{nm}(\x) G^{>,\l}_{m'n'}(\x-\o) 
\end{align}
The quantities that annihilate the $Q=0$ state such as local Green's function, are projected as
\begin{equation}
  G_{\a\b}(t,t') = \lim_{\l\to\infty} \frac{G^{\l}_{\a\b}(t,t')}{Q^{\l}}
\end{equation}
The retarded component of projected local Green's function is
\begin{align}
  G&^{R}_{\a\b}(\o) \notag \\
  =&  \lim_{\l\to\infty}\frac{1}{i\sum_{m} (-1)^{m} \int \frac{d\x}{2\p} G^{<,\l}_{mm}(\x)}
        i\sum_{mnm'n'} F^{\b*}_{mm'}F^{\a}_{n'n} \notag \\
        &\times\displaystyle \int \frac{d\x}{2\p}
          G^{R,\l}_{nm}(\x) G^{<,\l}_{m'n'}(\x-\o) 
         +G^{<,\l}_{nm}(\x) G^{A,\l}_{m'n'}(\x-\o)  \\
  =&  \lim_{\l\to\infty}\frac{1}{i\sum_{m} (-1)^{m} \int \frac{d\x}{2\p} \z G^{<}_{mm}(\x)}
        i\sum_{mnm'n'} F^{\b*}_{mm'}F^{\a}_{n'n} \notag \\
        &\times\displaystyle\int \frac{d\x}{2\p}
          G^{R}_{nm}(\x) \z G^{<}_{m'n'}(\x-\o) 
         +\z G^{<}_{nm}(\x) G^{A}_{m'n'}(\x-\o) \\
  =&  \frac{i}{Q}
        \sum_{mnm'n'} F^{\b*}_{mm'}F^{\a}_{n'n} \notag \\
        &\times\int \frac{d\x}{2\p}
          G^{R}_{nm}(\x) G^{<}_{m'n'}(\x-\o) 
         +G^{<}_{nm}(\x) G^{A}_{m'n'}(\x-\o) \label{eq:GRab}
\end{align}
where the projected pseudoparticle number, 
\begin{equation}
  Q = i\sum_{m} (-1)^{m} \int \frac{d\x}{2\p} G^{<}_{mm}(\x). \label{eq:Q}
\end{equation}
Likewise, the lesser component is
\begin{equation}
  G^{<}_{\a\b}(\o) =
    \frac{i}{Q}\sum_{mnm'n'} F^{\b*}_{mm'}F^{\a}_{n'n} \int \frac{d\x}{2\p}
    G^{<}_{nm}(\x) G^{>}_{m'n'}(\x-\o) \label{eq:G<ab}
\end{equation}

\section{Computational Results}\label{result}
\subsection{Phase diagram}

\begin{figure}[t]
  \centering
  \includegraphics[width=0.8\linewidth]{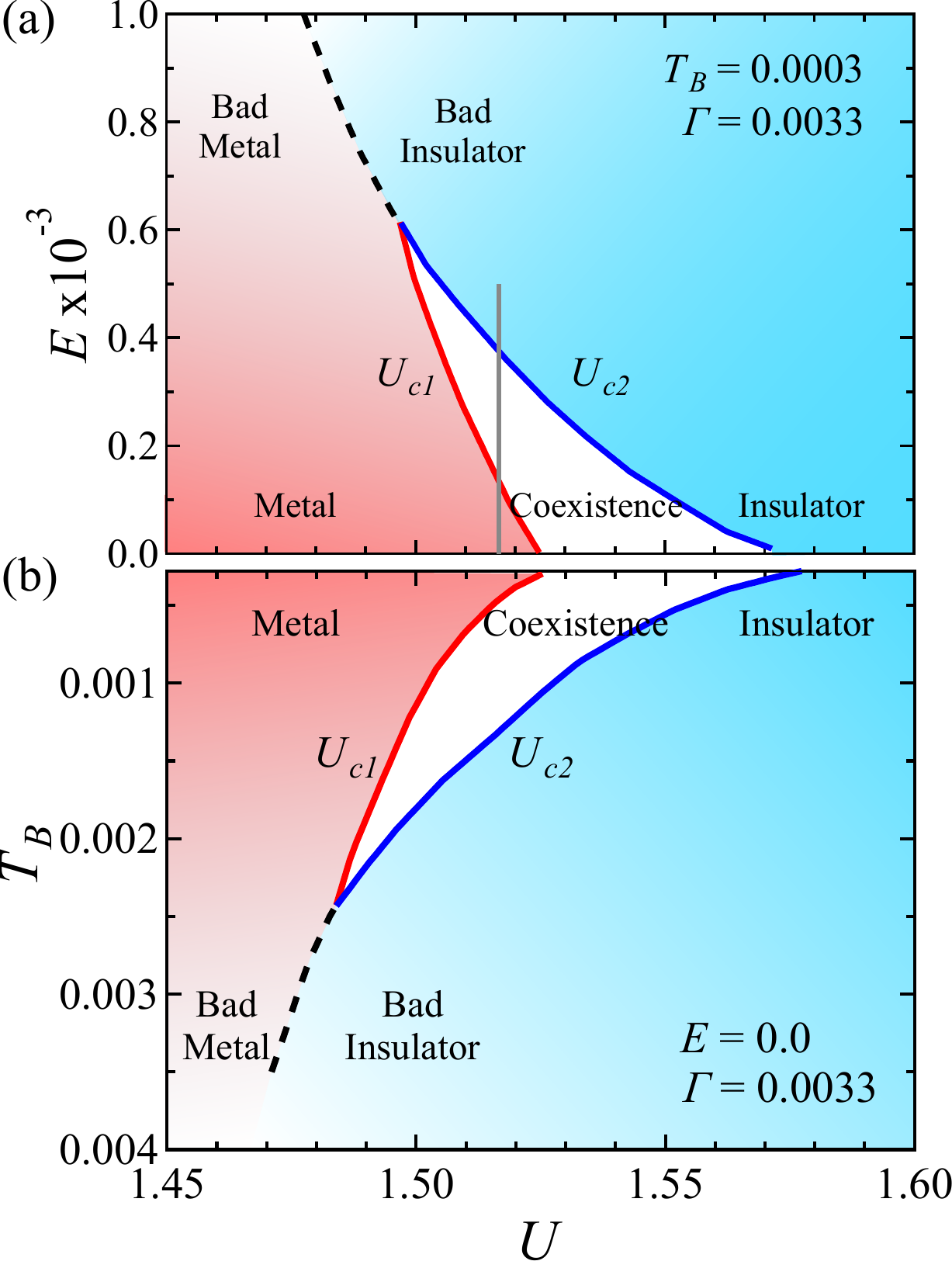}
  \caption{
  (a) The $U$ vs $E$ phase diagram with $T_{B} = 0.0003$, and $\Gamma = 0.0033$.
  A phase coexistence region is clearly bordered with $U_{c1}$ and $U_{c2}$ where each critical lines, $U_{c}$, are obtained by decreasing and increasing $E$, respectively.
  Above $E = 0.0006$, any clear first-order transition is not observed.
  The grey line indicates $U=1.517$, which is drawn for the spectral function and the self-energy discussions along the line.
  (b) The $U$ vs $T_{B}$ phase diagram with $E = 0$, and $\Gamma = 0.0033$. 
  This phase diagram belongs effectively to in-equilibrium ones.
  The phase diagram reproduced the well known single-band Hubbard model result.
  The critical lines, $U_{c}$, are obtained by scanning $U$.}
  \label{fig:pd}
\end{figure}
  
We test our non-equilibrium steady-state DMFT on the half-filled single-band dissipative Hubbard model under static and uniform electric field on three-dimensional cubic lattice.
For convenience, every energy scale is written in units of the half band width $D = 1$.
Fig.~\ref{fig:pd} (a), shows the Coulomb interaction strength, $U$ vs.a the electric field, $E$ phase diagram with the heat reservoir temperature, $T_{B} = 0.0003$, and the dissipation rate, $\Gamma = 0.0033$.
By increasing and decreasing $E$ for given $U$, we identify the critical $U_c$ lines.
Through these lines the system shows the first-order transitions observed by the spectral weights at the Fermi energy.
In Fig.~\ref{fig:pd} (b), we present the $U$ vs $T_{B}$ phase diagram with $E = 0$, and $\Gamma = 0.0033$, which corresponds to the equilibrium phase diagram due to the $E = 0$ setting.
The phase diagram reproduces the well known single-band Hubbard model result.
By comparing those two phase diagrams, we observe a clear resemblance between the $U$-$E$ and $U$-$T_{B}$ diagrams of the system  which are in contrast to the previously reported IPT results~\cite{Li2015}.
The difference in the result is attributed to the different choices of the impurity solvers implemented.
In equilibrium, the weak coupling second-order perturbation theory, commonly known as IPT, reproduces the correct strong coupling limit~\cite{Georges1996}.
This is, however, not the case when the system is out-of-equilibrium~\cite{Aoki2014}.
We use the strong coupling perturbation expansion that is conserved, albeit truncated at the lowest order (NCA), to investigate the strong coupling regime of the system.
We argue that the result in Fig.~\ref{fig:pd}(a) would describe the model more accurately.

\subsection{Metal-to-Insulator Transition (MIT) under electric field}

\begin{figure*}
  \centering
  \begin{minipage}[t]{.49\textwidth}
    \includegraphics[width=\linewidth]{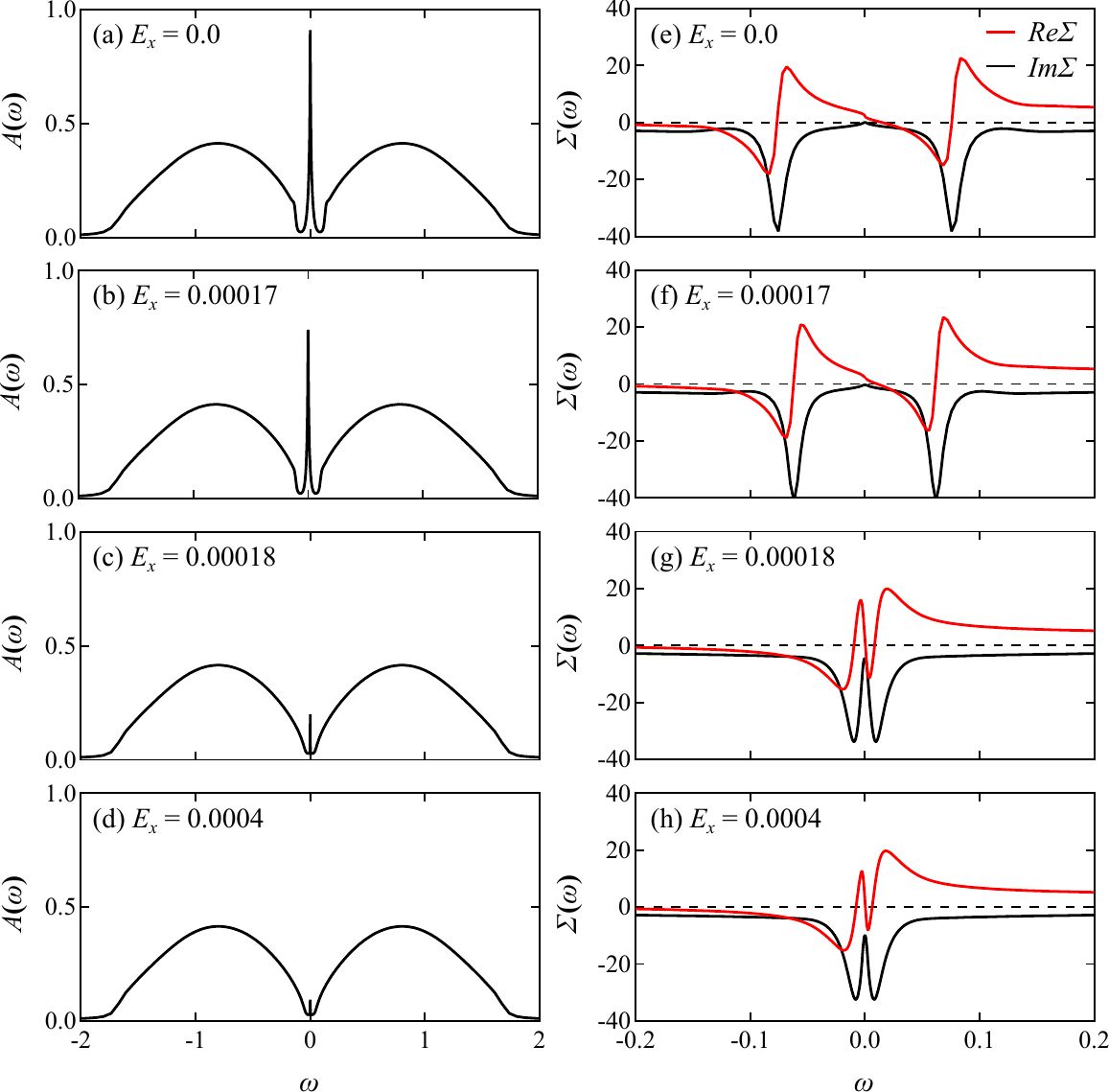}
    \caption{(a-d) Local spectral weight, $A(\o)$ at $U = 1.517$, $\Gamma = 0.0033$, $T_{B} = 0.0003$, for $E =$ 0, 0.00017, 0.00018, and 0.0004.
    Until $E=0.00017$ the spectral function shows strong coherent peak at the Fermi level.
    As $E$ is increased slightly to 0.00018 (c), the resonance peak is suppressed and the system becomes insulating.
    (e-h) Real and imaginary part of the retarded component of the local interaction self-energy $\S^{R}_{U}(\o)$ near the Fermi level at $U = 1.517$, $\Gamma = 0.0033$, and $T_{B} = 0.0003$, for $E =$ 0, 0.00017, 0.00018, and 0.0004.}
    \label{fig:alongE}
  \end{minipage}\hfill
  \begin{minipage}[t]{.49\textwidth}
    \includegraphics[width=\linewidth]{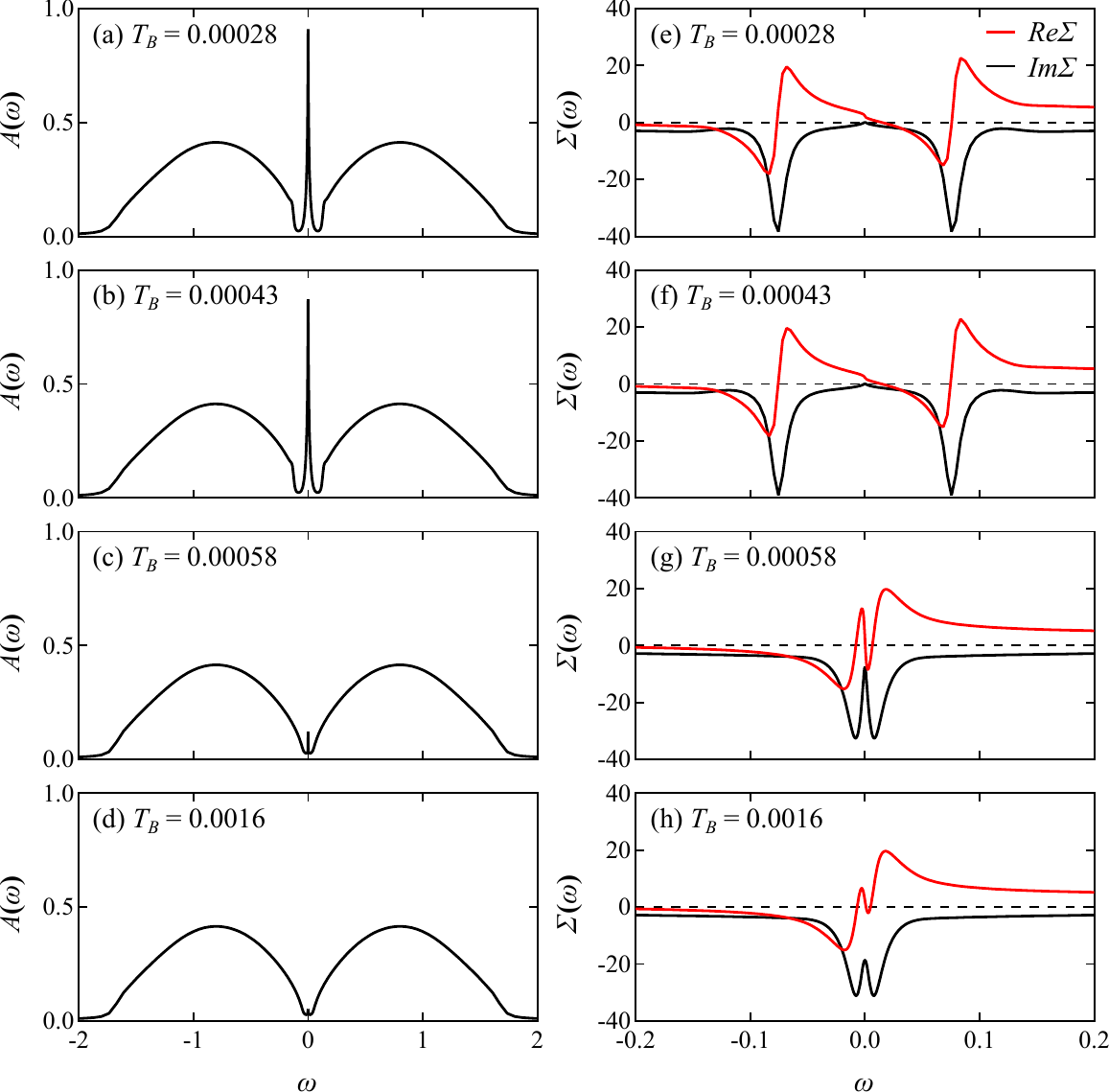}
    \caption{(a-d) Local spectral weight, $A(\o)$ at $U = 1.517$, $\Gamma = 0.0033$, $E = 0$, for $T_{B} =$ 0.00028, 0.00043, 0.00058, and 0.0016.
    Until $T_{B}=0.00043$ the spectral function shows strong coherent peak at the Fermi level.
    As $T_{B}$ is increased slightly (c), spectral weight at the Fermi level is suppressed and the system undergoes phase transition.
    (e-h) Real and imaginary part of the retarded component of the local interaction self-energy $\S^{R}_{U}(\o)$ near the Fermi level at $U = 1.517$, $\Gamma = 0.0033$, $E = 0$, for the same $T_{B}$ values.  }
    \label{fig:alongT}
  \end{minipage}
\end{figure*}

\begin{figure}[t]
  \centering
  \includegraphics[width=\linewidth]{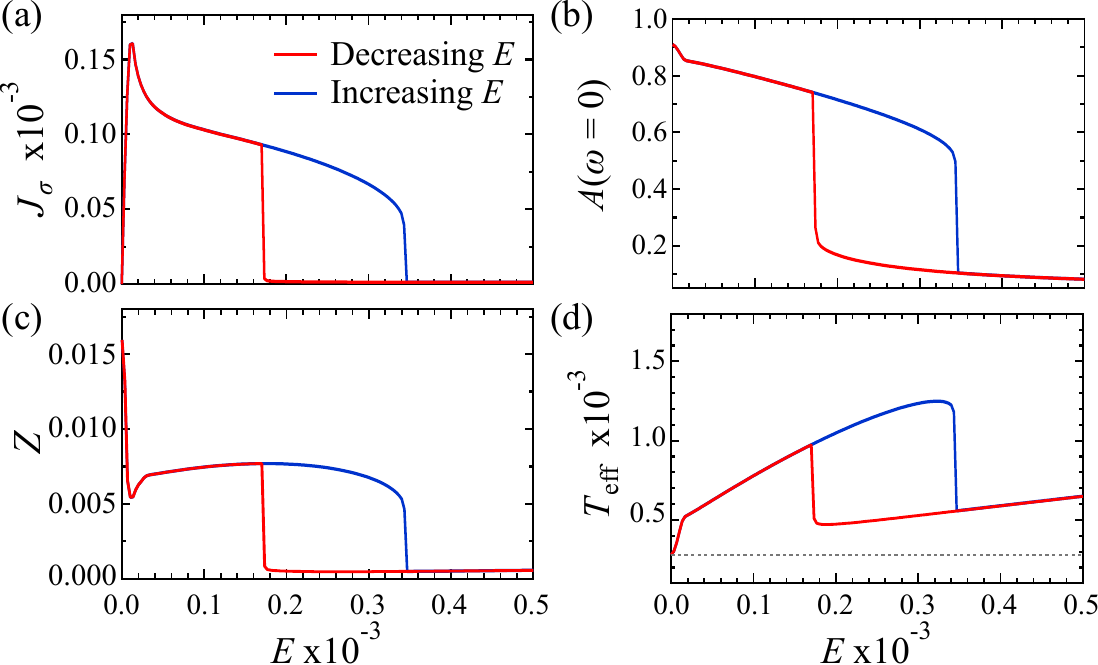}
  \caption{(a) The current per spin, $J_\s$, 
  (b) The the spectral weight at the Fermi level, $A(\o=0)$, 
  (c) The quasiparticle weights, $Z$ and 
  (d) The effective temperatures, $T_{\text{eff}}$ with respect to the $E$
  at $U = 1.517$, $\Gamma = 0.0033$, and $T_{B} = 0.0003$ (grey line in Fig.~\ref{fig:pd} (a)).
  First by increasing $E$ (blue line) then decreasing $E$ (red line), the electric field induces metal-to-insulator hysteresis.
  The grey dotted line in (d) indicates $T_{B}$.
  The system deviates from the linear response region very early, $E < 0.02\times10^{-3}$.}
  \label{fig:p4}
\end{figure}

To study the similarity between the $U$-$E$ and $U$-$T_{B}$ diagrams further, we compare the spectral functions and the local self energies across the first critical line $U_{c1}$ (along the solid grey line shown in Fig.~\ref{fig:pd}(a)).
Fig.~\ref{fig:alongE}(a)-(d) show the local spectral weights, $A(\o)$, and Fig.~\ref{fig:alongE}(e)-(h) show the real and imaginary parts of the retarded component of the local interaction self energies $\S^{R}_{U}(\o)$ near the Fermi level at $U = 1.517$, $\Gamma = 0.0033$, and $T_{B} = 0.0003$, for $E =$ 0, 0.00017, 0.00018, and 0.0004, respectively.
As the electric field $E$ is increased from zero, the coherent peak at $E_{F}$ is clearly observed and robust against the initial development of $E$.
And when the $E$ is around $0.00018$, the coherent peak abruptly vanishes and the system becomes insulating.
Figure~\ref{fig:alongT}(a)-(d) present, as displayed in Fig.~\ref{fig:alongE}, the spectral functions, $A(\o)$, and Fig.~\ref{fig:alongT}(e)-(h) show the local self energies 
$\S^{R}_{U}(\o)$ at $U = 1.517$, $\Gamma = 0.0033$, $E = 0$, for $T_{B} =$ 0.00028, 0.00043, 0.00058, and 0.0016, respectively.
Upon increasing $T_{B}$, around $T_{B} = 0.0005$ the first-order transition from metallic to insulating phase is observed.
Comparing Fig.~\ref{fig:alongE}(a)-(d) with Fig.~\ref{fig:alongT}(a)-(d), we observe that at the given heat bath temperature, the electric field-induced spectral function evolution is effectively the same as the $T$-evolution of the spectral weights in equilibrium.
The three-peak structure in low $E$ (or $T_B$), the abrupt collapse of the coherent peak in the three-peak structure indicating the first-order transition, and the Mott insulating density of states consisting purely of Hubbard bands are well shown.
In the spectral functions obtained by IPT, each Hubbard band on both sides is split into two respectively, making it look like a five-peaked structure.~\cite{Li2015}.

The retarded component of the local interaction self energies evolves likewise, hence given qualitatively similar quasiparticle weights, and the scattering rates.
The local distribution function,
\begin{equation}
  f_{\text{loc}}(\o) = -\frac{\Im G^{<}(\o)}{2\Im G^{R}(\o)} = \frac{1}{e^{\o/T_{\text{eff}}} + 1}
\end{equation}
 corresponds to the Fermi Dirac distribution if the system is in equilibrium, and evolves to a distribution with high effective temperature, $T_{\text{eff}}$.
This effective temperature can be fitted to the Fermi-Dirac distribution.
In the given model, the local distributions are close to the Fermi-Dirac function.

In Fig.~\ref{fig:p4}, we sweep the electric field from zero to 0.0005 then back to zero in the vicinity of Mott transition, along the grey line in Fig.~\ref{fig:pd}, using the self-consistent solution from the previous calculation.
Figure~\ref{fig:p4}(a) shows the current per spin, $J_\s$, or the $I-V$ curve.
The current is calculated using the lesser Green's functions as shown below,
\begin{equation}
  J_\s = t \Re[G^{<}_{0,1}(0) + G^{<}_{-1,0}(0)].
\end{equation}
A linear response regime appears but exists in a very limited region below $E < 0.02\times10^{-3}$.
So the system generally shows a drastic nonlinear behavior for all the quantities we measure.
Especially, metal-to-insulator transition and its hysteresis are notable along with the nonlinearity.
The $E$-$J_\s$ curve shows a sharp peak at the end of the linear response regime, which is absent in the IPT result.~\cite{Li2015}
Due to this sharp peak structure, the linear and the nonlinear regions are easily distinguishable.
The MIT is clearly visible not only in the evolution of the current above but also in the evolution of the spectral weight at E$_F$ in Fig.~\ref{fig:p4}(b).
Since the system is near the critical point, the quasiparticle weight, $Z = [1-\frac{d\Re\S(\o)}{d\o}|_0]^{-1}$, is strongly renormalized but shows the hysteresis as well in Fig.~\ref{fig:p4}(c).
In Fig.~\ref{fig:p4}(d), we show the hysteresis of the effective temperature.
At zero field, the system is in equilibrium having the same temperature as the reservoir, $T_{\text{eff}} = T_B$ (the grey dotted line).
Then as the electric field is increased, the $T_{\text{eff}}$ rises a magnitude of order higher than the environment, and then abruptly cools down when the system becomes an insulator.
Recently using rare-earth doped fluorescent material, a measurement of the local temperature of a sample undergoing MIT by dc voltage is devised~\cite{Zimmers:2013aa}.
We expect for a system where one band Hubbard model is applicable, such as \ce{NiO}, and \ce{Cr} doped \ce{V_2 O_3}, one can observe similar nonlinear behavior of the local temperature as in Fig.~\ref{fig:p4}(d).

\subsection{Heat dissipation rate $\Gamma$-dependent phase diagrams : the importance of the geometry of a system}

From Eq.~\ref{eq:action} and Eq.~\ref{eq:action2}, the heat dissipation, $\S_{\Gamma}(\o)$ is included in the hybridization function.
In the non-equilibrium steady-state state, the correlated fermions have the hybridizations not only to themselves but also to the attached free fermion bath.
In the aspect of the hybridization function, the non-equilibrium hybridization function is just a new hybridization function rescaled by the fermionic bath, $\Gamma$ from an equilibrium hybridization function.
So, the observed similarity between the $U$-$T$ and the $U$-$E$ diagrams is well understood by the similar form of the hybridization functions between the equilibrium state influenced by the temperature and the non-equilibrium state influenced by the electric field and the heat bath fermions.
The electric field is a source that rises the system's temperature, and the fermionic bath is a source that drains it.
The effective temperature $T_{\text{eff}}$ in the non-equilibrium steady state is a result of this balance.
In this sense, we can consider $T_{\text{eff}}$ as a new temperature which is rescaled from the equilibrium thermal energy inside the system after the competition between the electric field and the fermionic heat bath.
The observed similarity between the $U$-$T$ and the $U$-$E$ diagrams can be understood by considering the effective temperature $T_{\text{eff}}$ as that of equilibrium state.

\begin{figure}[t]
  \centering
  \includegraphics[width=0.8\linewidth]{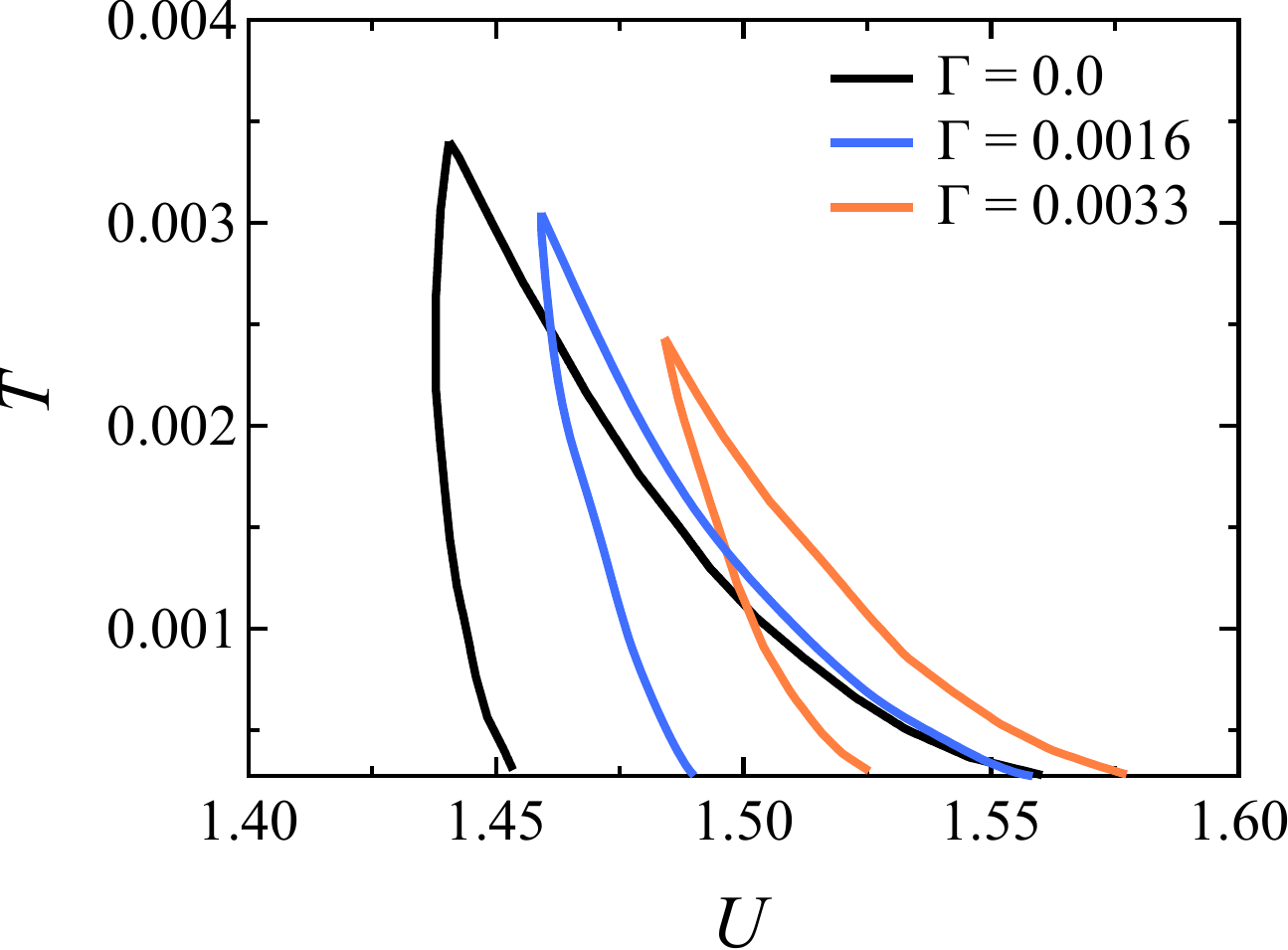}
  \caption{Equilibrium ($E = 0$) $U$ v.s. $T$ phase diagrams with $\Gamma = 0$ (black), $\Gamma = 0.0016$ (blue), and $\Gamma = 0.0033$ (orange).
  As the dissipation rate, $\Gamma$, increases the system shows smaller coexistence region, and the critical $U$ becomes larger.
  The $U_{c1}$ is more affected, and system becomes more coherent at given $U$ with larger $\Gamma$.}
  \label{fig:Gpd}
\end{figure}

In response to the varying $\Gamma$s, the new temperature scale, $T_{\text{eff}}$ leads to the variation of the $U$-$T$ phase diagrams as seen in Fig.~\ref{fig:Gpd}.
The sensitive change of the phase diagram is noticeable with respect to the $\Gamma$s.
When the $\Gamma$ increases, the system tends to have smaller coexistence region. 
And the critical $U$ becomes larger making the system to prefer a more coherent state. 
With a larger $\Gamma$, the state becomes more resistant to the applied electric field and the system prefers to be metallic at the larger $U$.
These different MIT phase diagrams with respect to the different heat dissipation rates $\Gamma$s lead us reach a very important conclusion especially when we think of how the heat dissipation rates $\Gamma$s will be changed.

This heat dissipation dependence on the criticality can be manifested through the difference in the sample's surface morphology, which is indeed observed in an experiment~\cite{Zimmers:2013aa}.
The result showed that the metallic \ce{VO_2} with a larger surface area is more resistant to the applied electric field than the sample with a smaller surface area so that the transition to the insulating \ce{VO_2} occurred in higher electric field in case of the sample with a larger surface area.
In this reason, the sample's morphology is a very important factor that determines the electronic structure of a given correlated system under the electric field.

\section{Conclusion}\label{conclude}

In summary, we investigate the half-filled single-band dissipative Hubbard model under electric field using the non-equilibrium steady-state dynamical mean field theory by means of the non-crossing approximation impurity solver.
Both the electric field and the Coulomb interaction are treated on the equal footing.
By building the dissipative Hubbard model where the additional energy acquired from the field is dissipated by the fictitious free Fermion reservoir attached to each correlated electrons, and calculating physical quantities such as current induced by the electric field, local distribution functions which gives effective local temperature we observe the electric field induced metal-to-insulator transition and the phase coexistence region in the phase diagram.
We argue that the electric field near the criticality takes the role of thermal energy in equilibrium and the metal-to-insulator transition by the field is qualitatively similar as in equilibrium MIT by using more sophisticated impurity solver.
We expect these results will describe correctly the behavior of compounds such as \ce{NiO}, \ce{Cr} doped \ce{V_2 O_3}.
Furthermore, the formalism given here is relatively easy to extend to EDMFT, where the bosonic degree of freedom, such as phonon, is accounted for hence giving more accurate electric current description of the system.
It should also be possible to incorporate the formalism with the density functional theory.

\begin{acknowledgments}
We thank Gabriel Kotliar, Kristjan Haule, Jong Han, and Seung-Sup Lee for fruitful discussions.
This work was supported by the Global-LAMP Program of the National Research Foundation of Korea (NRF) grant funded by the Ministry of Education (No.~RS-2023-00301976).
\end{acknowledgments}
 
\bibliography{references}

\onecolumngrid
\appendix

\section{NESS-DMFT equations}

Construction of the NESS-DMFT equations were carried out by Li \etal~\cite{Li2015}. Here we do similar derivation for the general multi-orbital case. 
The $ll'$ component of the retarded Green's function (see Eq.~\ref{eq:GRdef}) is an $N \times N$ block matrix $\mathbf{g}_{ll'} \equiv [\mathbf{G}^{R}(k_\perp, \o)]_{ll'}$ and by the defining equation of inverse,
\begin{equation}\label{eq:Ginversion}
  \mathbf{g}_{0l-1}\mathbf{t} + \mathbf{g}_{0l}\mathbf{A}_{l} + \mathbf{g}_{0l+1}\mathbf{t} = \mathbf{I}\d_{0l}.
\end{equation}
For $l = 0$,
\begin{equation}\label{eq:GR00k}
  \mathbf{g}^{-1}_{00}
  = \mathbf{A}_{0} + \left[\mathbf{g}^{-1}_{00}\mathbf{g}_{0-1} + \mathbf{g}^{-1}_{00}\mathbf{g}_{01}\right]\mathbf{t}
\end{equation}
and by recursively applying Eq.~\ref{eq:Ginversion} with $l\neq0$, the last two terms can be expressed as
\begin{equation}\label{eq:FR}
  \mathbf{g}^{-1}_{00}\mathbf{g}_{0\pm1} = -\mathbf{t} 
    \left[\mathbf{A}_{\pm1}-\mathbf{t}\left[\mathbf{A}_{\pm2}-\mathbf{t}\left[\mathbf{A}_{\pm3}-\cdots
    \mathbf{t}\right]^{-1}\mathbf{t}\right]^{-1}\mathbf{t}\right]^{-1}.
\end{equation}
We consider the right and left chains separately for later use.
\begin{equation}
  \left[\mathcal{F}^{R}_{\pm}(k_\perp, \o)\right]^{-1} =
    \begin{bmatrix}
      \mathbf{A}_{\pm1} & \mathbf{t} & & \\
      \mathbf{t} & \mathbf{A}_{\pm2} & \mathbf{t} & \\
      & \mathbf{t} & \mathbf{A}_{\pm3} & \ddots \\
      & & \ddots & \ddots \\
    \end{bmatrix}
\end{equation}
Following similar steps ($\mathbf{f}^{\pm}_{ll'} \equiv [\bm{\mathcal{F}}^{R}_{\pm}(k_\perp, \o)]_{ll'}$),
\begin{align}
  1 &= \mathbf{f}^{\pm}_{11}\mathbf{A}_{\pm1} + \mathbf{f}^{\pm}_{12}\mathbf{t},
  \label{eq:Finversion1}\\
  0 &= \mathbf{f}^{\pm}_{1l-1}\mathbf{t}+\mathbf{f}^{\pm}_{1l}\mathbf{A}_{\pm l}+\mathbf{f}^{\pm}_{1l+1}\mathbf{t},
  \label{eq:Finversion2}
\end{align}
\begin{equation}\label{eq:FRdef2}
  \mathbf{f}^{\pm}_{11} =
    \left[\mathbf{A}_{\pm1}-\mathbf{t}\left[\mathbf{A}_{\pm2}-\mathbf{t}\left[\mathbf{A}_{\pm3}-\cdots
    \mathbf{t}\right]^{-1}\mathbf{t}\right]^{-1}\mathbf{t}\right]^{-1}
    \equiv \mathbf{F}^{R}_{\pm}(k_\perp, \o-\ph_{\pm1}).
\end{equation}
Thus Eq.~\ref{eq:FR} can be written in terms of $\mathbf{f}^{\pm}_{11}$,
\begin{equation}\label{eq:F11}
  \mathbf{g}^{-1}_{00}\mathbf{g}_{0\pm1} = -\mathbf{t}\mathbf{f}^{\pm}_{11}
\end{equation}
and self-similarity of $\mathbf{F}$ leads to
\begin{equation}\label{eq:FRrecur}
  \mathbf{F}^{R}_{\pm}(k_\perp, \o-\ph_{\pm1}) =
    \left[\mathbf{A}_{\pm1}-\mathbf{t} \mathbf{F}^{R}_{\pm}(k_\perp, \o-\ph_{\pm2}) \mathbf{t}\right]^{-1}.
\end{equation}
Finally, the retarded component of the local Green's function is from Eq.~\ref{eq:GR00k}, ~\ref{eq:F11}, and ~\ref{eq:FRdef2},
\begin{align}
  \mathbf{G}^{R}(\o) = \sum_{k_\perp}
  \left[\mathbf{G}^{R}(k_\perp, \o)_{00}\right]^{-1}
  = \sum_{k_\perp} \big[\mathbf{A}_{0} - \mathbf{t} \big( \mathbf{F}^{R}_{+}(k_\perp, \o-\ph_1) + \mathbf{F}^{R}_{-}(k_\perp, \o-\ph_{-1}) \big) \mathbf{t} \big]^{-1}
\end{align}

From the Dyson's equation for the lesser component of the local Green's function (Eq.~\ref{eq:DysonG<}), the $00$ component is
\begin{align}\label{eq:dysonL}
  \left[\mathbf{G}^{<}(k_\perp,\o)\right]_{00} =&
    \mathbf{G}^{R}(k_\perp,\o)_{00} \bm{\S}^{<}_{\text{tot}}(\o)_{0} \mathbf{G}^{A}(k_\perp,\o)_{00}
   +\sum_{\substack{p<0\\p>0}} \mathbf{G}^{R}(k_\perp,\o)_{0p} \bm{\S}^{<}_{\text{tot}}(\o)_{p}
      \mathbf{G}^{A}(k_\perp,\o)_{p0} \notag \\
  \mathbf{g}^{<}_{00} =&
    \mathbf{g}_{00}\bm{\S}^{<}_{\text{tot}}(\o)_{0}\mathbf{g}^{\+}_{00}
    +\sum_{p>0} \mathbf{g}_{0\pm p} \bm{\S}^{<}_{\text{tot}}(\o)_{\pm p} \mathbf{g}^{\+}_{\pm p0}.
\end{align}
Similarly, it is convenient to think of right/left chains.
From Eq.~\ref{eq:Ginversion}, Eq.~\ref{eq:Finversion1} with $l=1$, using Eq.~\ref{eq:F11}, we get
$\mathbf{g}^{-1}_{00}\mathbf{g}_{0\pm2} = -\mathbf{t}\mathbf{f}^{\pm}_{12}$. With $l=2$ we get
$\mathbf{g}^{-1}_{00}\mathbf{g}_{0\pm3} = -\mathbf{t}\mathbf{f}^{\pm}_{13}$ and so on. Therefore,
\begin{equation}\label{eq:F1p}
  \mathbf{g}_{0\pm p} = -\mathbf{g}_{00}\mathbf{t}\mathbf{f}^{\pm}_{1p}
\end{equation}
With the Dyson's equation for the $\mathbf{F}^{<}$, Eq.~\ref{eq:dysonL} becomes,
\begin{equation}\label{eq:G<00k}
  \mathbf{g}^{<}_{00} =
    \mathbf{g}_{00} \left[\bm{\S}^{<}_{\text{tot}}(\o)_{0}
     +\mathbf{t}\left(\mathbf{f}^{<,+}_{11} + \mathbf{f}^{<,-}_{11} \right) \mathbf{t}
     \right] \mathbf{g}^{\+}_{00}.
\end{equation}
Doing the same calculations for a right/left semi infinite chains starting with $l=2$ (defining $\mathbf{h}$), and realizing the self similarity to $l=1$ ($\mathbf{f}$), we can write $\mathbf{f}^{<}_{11}$ as
\begin{align}\label{eq:F<recur}
  \mathbf{f}^{<,\pm}_{11} &\equiv \mathbf{F}^{<}_{\pm}(k_\perp,\o-\ph_{\pm1}) = 
    \mathbf{f}^{\pm}_{11}\left[
      \bm{\S}^{<}_{\text{tot}}(\o)_{\pm1} +\mathbf{t}
      \left( \sum_p \mathbf{h}^{\pm}_{2p} \bm{\S}^{<}_{\text{tot}}(\o)_{p} \mathbf{h}^{\pm,\+}_{p2} \right)
    \mathbf{t} \right] \mathbf{f}^{\pm,\+}_{11} \notag \\
  &= \mathbf{F}^{R}_{\pm}(k_\perp,\o-\ph_{\pm1}) \left[
      \bm{\S}^{<}_{\text{tot}}(\o-\ph_{\pm1}) +\mathbf{t} \mathbf{F}^{<}_{\pm}(k_\perp,\o-\ph_{\pm2}) \mathbf{t}
    \right] \mathbf{F}^{A}_{\pm}(k_\perp,\o-\ph_{\pm1})
\end{align}
Thus the lesser component of the local Green's function is, from Eq.~\ref{eq:G<00k} and ~\ref{eq:F<recur},
  \begin{align}
    \mathbf{G}^{<}(\o)
    = \sum_{k_\perp} \mathbf{G}^{R}(k_\perp, \o)_{00}
      \left[
        \bm{\S}^{<}_{\text{tot}}(\o) + \mathbf{t}
        \left(
          \mathbf{F}^{<}_{+}(k_\perp,\o-\ph_{1}) +\mathbf{F}^{<}_{-}(k_\perp,\o-\ph_{-1})
        \right)
        \mathbf{t}
      \right] \mathbf{G}^{A}(k_\perp, \o)_{00}
  \end{align}

\section{NESS-NCA numerical implementations}

\subsubsection{Imaginary parts}

Every quantity in SSNCA self-consistent equations can be expressed only with its imaginary part denoted
\begin{equation}
  \overline{F}^{X}(\o) \equiv \Im[F^{X}(\o)]
\end{equation}
where $X\in\{R,<,>,\ldots\}$.
The real part of retarded components is obtained by the Kramers-Kronig relation.
In this way we can avoid storing complex numbers in the memory and complex number arithmetics which are computationally heavier than the ordinary floating number calculations.
\begin{equation}
  \Re[F^{R}(\o)] = \frac{1}{\p}\mathcal{P}\int^{\infty}_{-\infty} d\x \frac{\overline{F}^{R}(\x)}{\x-\o}
\end{equation}
And by the symmetry of the contour-ordered Green's function, i.e. $[G^{<>}(t)]^* = -G^{<>}(-t)$ $\Rightarrow$ $[G^{<>}(\o)]^* = -G^{<>}(\o)$, the lesser(greater) component is purely imaginary.
\begin{equation}
  F^{<>}(\o) = i \overline{F}^{<>}(\o)
\end{equation}
Also note that $2\overline{F}^{R}(\o) = \overline{F}^{>}(\o) - \overline{F}^{<}(\o)$ and for the projected quantities, $2\overline{F}^{R}(\o) = \overline{F}^{>}(\o)$.
Furthermore factor $(-1)^{mn}$  can be absorbed in the lesser components
\begin{equation}
  \overline{F}^{\le}_{mn}(\o) \equiv (-1)^{mn} \overline{F}^{<}_{mn}(\o)
\end{equation}
In this notation, the pseudoparticle self-energy, pseudoparticle number, and the local Green's function becomes
  \begin{align}
    \overline{\S}^{R\le}_{mn}(\o) &=
      -\sum_{\a\b m'n'}\int\frac{d\x}{2\p} \overline{G}^{R\le}_{m'n'}(\x)
      \Big[
         F^{\b}_{mm'}F^{\a*}_{n'n} \overline{\D}^{\le>}_{\a\b}(\x-\o)
        +F^{\b*}_{mm'}F^{\a}_{n'n} \overline{\D}^{>\le}_{\b\a}(\o-\x)
      \Big]
      \label{eq:Smn} \\
    Q &= -\sum_{m} \int \frac{d\x}{2\p} \overline{G}^{\le}_{mm}(\x) \label{eq:Q2}\\
    \overline{G}^{R<}_{\a\b}(\o)
    &=  \frac{(-1)^{1,0}}{Q}
          \sum_{mnm'n'} F^{\b*}_{mm'}F^{\a}_{n'n} \int \frac{d\x}{2\p}
          \Big[
            \overline{G}^{R  \le}_{nm}(\x) \overline{G}^{\le R}_{m'n'}(\x-\o) 
           +\overline{G}^{\le\le}_{nm}(\x) \overline{G}^{R   R}_{m'n'}(\x-\o) 
          \Big]
      \label{eq:Gab}
  \end{align}
  in Eq.~\ref{eq:Gab}, we used the fact that $G_{nm}$ is fermionic, $G_{m'n'}$ is bosonic, and
  $\overline{G}^{>}_{m'n'}(\o) = 2\overline{G}^{R}_{m'n'}(\o) =-2\overline{G}^{A}_{m'n'}(\o)$.

\subsubsection{Kramers-Kronig relation}

  When both real and imaginary parts of a function is required, such as for solving a Dyson's equation, we can use the Kramers-Kornig relation.
  Let $F(z) = F'(z) + iF''(z)$, $F:\mathbb{C}\to\mathbb{C}$, where single(double) prime indicates real(imaginary) part.
  Suppose $F$ is analytic in the closed upper half-plane of $z$ and vanishes like $1/|z|$ or faster as $|z|\to\infty$, then
  \begin{align}
    F'(z)  &=  \frac{1}{\p}\mathcal{P}\int_{-\infty}^{\infty}dz' \frac{F''(z')}{z'-z} \\
    F''(z) &= -\frac{1}{\p}\mathcal{P}\int_{-\infty}^{\infty}dz' \frac{F' (z')}{z'-z}
  \end{align}
  where $\mathcal{P}$ denotes the Cauchy principle value.
  Thus the full function $F$ can be constructed with only one of its part.

  The implementation of the relation is given by
  \begin{align*}
    \p F'(x_0)
    &= \lim_{\e\to0^{+}}\left(\int_{-W}^{x_0-\e}+\int_{x_0+\e}^{W}\right) dx\frac{F''(x)}{x-x_0} \\
    &= \lim_{\e\to0^{+}}\left(\int_{-W}^{x_0-\e}+\int_{x_0+\e}^{W}\right) dx\frac{F''(x)-F''(x_0)}{x-x_0}
       +\frac{F''(x_0)}{x-x_0} \\
    &= \lim_{\e\to0^{+}}\left(\int_{-W}^{x_0-\e}+\int_{x_0+\e}^{W}\right) dx\frac{F''(x)-F''(x_0)}{x-x_0} +F''(x_0)\left(\int_{-W}^{x_0-\e}+\int_{x_0+\e}^{W}\right) \frac{dx}{x-x_0} \\
    &= \lim_{\e\to0^{+}}\left(\int_{-W}^{x_0-\e}+\int_{x_0+\e}^{W}\right) dx\frac{F''(x)-F''(x_0)}{x-x_0} +F''(x_0) \ln \left[\frac{x_0-\e-x_0}{-W-x_0}\frac{W-x_0}{\e}\right] \\
  \end{align*}
  \begin{align*}
    \p F'(x_i)
    \Rightarrow&
       \sum_{j \ne i}\D x\frac{F''(x_j)-F''(x_i)}{x_j-x_i}+\frac{\D x}{2}\left[\frac{F''(x_i)-F''(x_{i-1})}{x_i-x_{i-1}}+\frac{F''(x_{i+1})-F''(x_i)}{x_{i+1}-x_i}\right] +F''(x_i) \ln \left[\frac{W-x_0}{-(-W)+x_0}\right]
  \end{align*}
  For the numerical stability the divergent term is subtracted and calculated analytically and the second term in the last equation handles the singular $(j=i)$ case by averaging values at $i\pm 1$.
  $\pm W$ is chosen such that $F''(\pm W)$ is negligible.
  In this way the function being integrated is smooth everywhere.

\subsubsection{Subtraction of Lorentzian distribution}
  \label{sub:Subtraction of Lorentzian distribution}

  Often the pseudoparticle Green's functions are sharply peaked at some energy, and this causes problem when convolving with the hybridization functions.
  This can be avoided if one subtracts a Lorentzian distribution from the Green's function and treats the Lorentzian semi-analytically.
  Let us define,
  \begin{equation}
    L(\o) = P\frac{\g}{(\o-x)^2 + \g^2}, \qquad g(\o) = G(\o) - L(\o)
  \end{equation}
  where, $P$ is a constant depicting the size of $L$, $x, \g$ are the center and the width of $L$.
  There are four integrals to consider: $\int d\x L(\x)A(\x-\o), \int d\x A(\x)L(\x-\o), \int d\x L_1(\x)L_2(\x-\o),$ and $\int d\x L(\x)$.
  The last two are trivial
  \begin{align}
    \int d\x L(\x) &= P\p \\
    \int d\x L_1(\x)L_2(\x-\o) &= P_1 P_2 \p\frac{\g_1+\g_2}{(x_1 -x_2 -\o)^2 + (\g_1+\g_2)^2}
  \end{align}
  First two integrals are identical noting $\o \to -\o$.
  Dividing the integral into N slices and fiddling the terms we have,
  \begin{equation}
    P\sum_i \int_{\x_i}^{\x_{i+1}} d\x \frac{\g A(\x)}{(\x+\o-x_0)^2+\g^2} = P\sum_i R_i
  \end{equation}
  \begin{align*}
    R_i
    =&  \left[\tan^{-1}\left(\frac{\x+\o-x_0}{\g}\right)A(\x)\right]^{\x_{i+1}}_{\x_i}
        -\int_{\x_i}^{\x_{i+1}} d\x \tan^{-1}\left(\frac{\x+\o-x_0}{\g}\right)\frac{dA(\x)}{d\x} \\
    =&  \left[\tan^{-1}\left(\frac{\x+\o-x_0}{\g}\right)A(\x)\right]^{\x_{i+1}}_{\x_i}
        -\g \frac{\D A}{\D\x} \left[ \left(\frac{\x+\o-x_0}{\g}\right)\tan^{-1}\left(\frac{\x+\o-x_0}{\g}\right)
        -\frac{1}{2}\ln((\x+\o-x_0)^2+\g^2) \right]_{\x_i}^{\x_{i+1}} \\
    =&  \left(A(\x_i) - \frac{\D A}{\D\x}c_0\right)
        \left[\tan^{-1}\left(\frac{c_1}{\g}\right) - \tan^{-1}\left(\frac{c_0}{\g}\right)\right]
        +\frac{\D A}{\D\x}\frac{\g}{2}
        \left[\ln\left(\left(\frac{c_1}{\g}\right)^2+1\right)-\ln\left(\left(\frac{c_0}{\g}\right)^2+1\right)\right]
  \end{align*}
  where $c_{0,1}= \x_{i,i+1}+\o-x_0$.
  With 
  \begin{align*}
    \tan^{-1}(x) &\approx 
    \begin{cases}
      \displaystyle \text{sgn}(x)\frac{\p}{2}-\frac{1}{x}, & 1/x \to 0 \\
      \displaystyle x, & x \to 0 \\
    \end{cases} \\
    \frac{1}{2}\ln(x^2+1) &\approx
    \begin{cases}
      \displaystyle \ln(x) + \frac{1}{2x^2}, & 1/x \to 0 \\
      \displaystyle \frac{x^2}{2}, & x \to 0 \\
    \end{cases}
  \end{align*}
  one can get all the limiting cases except when $|\x_{i+1}-\x_{i}| \ll \g$.
  In this case the middle point value is enough
  \begin{equation}
    R_i = \g \frac{A(\x_{i+1})-A(\x_{i})}{2}\D\x \left[\left(\frac{\x_{i+1}-\x_{i}}{2}+\o-x_0 \right)^2+\g^2\right]^{-1}
  \end{equation}

  To determine the constants for a Lorentzian distribution, we express $G(\o)$ using the Taylor expansion around $z_0$ which makes $\o+\m-\Re\S(\o) = 0$, and using the perfect square expression
  \begin{align}
    G(\o) &\approx \frac{1}{A(\o-z_0)^2 + 2B(\o-z_0) + C} =  \frac{P\g}{(\o-x_0)^2 + \g^2} \\
      A &= \left.\frac{1}{2}\frac{d^2}{d\o^2}\frac{1}{G(\o)}\right|_{z_0},
      \qquad B = \left.\frac{1}{2}\frac{d}{d\o}\frac{1}{G(\o)}\right|_{z_0},
      \qquad C = \left.\frac{1}{G(\o)}\right|_{z_0} \notag\\
      x_0 &= z_0 - \frac{B}{A},
      \qquad \g^2 = \frac{C}{A} - \left(\frac{B}{A}\right)^2,
      \qquad P = \frac{1}{A\g} \notag
  \end{align}
  For example, with
  \begin{equation*}
    G(\o) = \frac{\widetilde{\S}(\o)}{[\o+\m-\Re\S(\o)]^2 + [\overline{\S}(\o)]^2}
  \end{equation*}
  we have, 
  \begin{align*}
    A =&
      \frac{1}{\widetilde{\S}(z_0)}
        \Bigg[(1-\Re\S'(z_0))^2 + \overline{\S}'(z_0)^2
              -2\overline{\S}(z_0) \overline{\S}'(z_0)\frac{\widetilde{\S}'(z_0)}{\widetilde{\S}(z_0)}
              + \overline{\S}(z_0)^2\left(\frac{\widetilde{\S}'(z_0)}{\widetilde{\S}(z_0)}\right)^2 
        \Bigg] \\
    B =& 
      \frac{1}{\widetilde{\S}(z_0)}
        \left[\overline{\S}(z_0) \overline{\S}'(z_0)
              -\frac{1}{2} \overline{\S}(z_0)^2\frac{\widetilde{\S}'(z_0)}{\widetilde{\S}(z_0)}
        \right] \\
    C =& 
      \frac{\overline{\S}(z_0)^2}{\widetilde{\S}(z_0)}
  \end{align*}

\subsubsection{Local gauge symmetry of pseudoparticles}
  \label{sub:Local gauge symmetry of pseudoparticles}

  The pseudo Green's functions is usually highly peaked around some energy $\o_0$.
  Therefore using non equi-distance mesh that has high density around $\o_0$ is a good strategy for both efficiency and accuracy.
  It is obvious that there is a local gauge symmetry $\o \to \o+\l$ for pseudoparticles.
  Hence one can move the whole pseudo quantities structure so that the maximum of a pseudo Green's function is centered around zero (User can select which pseudoparticle to use in the input file).
  The sufficient condition is
  \begin{equation}
    B(0,\l') = 0, \quad \text{ where } \quad
    B(\o,\l) \equiv \frac{d}{d\o}\left(\frac{1}{\overline{G}^{R}(\o;\l)}\right)
  \end{equation}
  $B(0,\l)$ is quadratic equation so if there is no solution it is sufficient to have $\l'$ that minimizes $B$.
  \begin{align*}
    B(\o,\l)
    =& \frac{d}{d\o}
        \left\{
          \frac{(\o-h+\l-\Re{\S^{R}}(\o))^2+(\overline{\S}^{R}(\o))^2}{\overline{\S}^{R}(\o)}
        \right\}
        \\
    =& -\frac{d\overline{\S}^{R}(\o)}{d\o}\frac{1}{(\overline{\S}^{R}(\o))^2}
          \Bigg\{
            (\l+\o-h-\Re{\S^{R}}(\o))^2 - (\overline{\S}^{R}(\o))^2 \\
        &   -2\left(1-\frac{d\Re{\S^{R}}(\o)}{d\o}\right)\overline{\S}^{R}(\o)
              \left(\frac{d\overline{\S}^{R}(\o)}{d\o}\right)^{-1} (\l+\o-h-\Re{\S^{R}(\o)})
          \Bigg\}
  \end{align*}
  Let $x = h+\Re{\S^{R}}(0)$,
  $b = -\left(1-\left.\frac{d\Re{\S^{R}}(\o)}{d\o}\right|_0\right)\overline{\S}^{R}(0)
      \left(\left.\frac{d\overline{\S}^{R}(\o)}{d\o}\right|_0\right)^{-1}$, and $c = -(\overline{\S}^{R}(\o))^2$, then
  \begin{equation}
    0 = (\l-x)^2 + 2b(\l-x) + c
  \end{equation}
  If $b^2-c < 0, \l = x-b$ which minimizes $B$ otherwise $\l = x - b \pm \sqrt{b^2-c}$ and one can choose $\l$ that maximizes $|\overline{G}^{R}(0;\l)|$.
  New $\l$ will be found at every iteration steps for the retarded Green's function, and the lesser Green's function will use $\l$ that is set by the retarded component.
\end{document}